\documentclass[journal, 10pt]{IEEEtran}
\usepackage{amssymb}
\usepackage{amsmath}
\usepackage{cite}
\usepackage{mathrsfs}
\usepackage{makecell}
\usepackage{url}
\usepackage{xcolor}
\usepackage{bbm}
\usepackage{cite,graphicx,amsmath,amssymb}
\usepackage{subcaption}
\usepackage{mdwtab}
 \usepackage{multirow}
\usepackage{fancyhdr}
\usepackage{hyperref}
\usepackage{amsthm}
\usepackage[linesnumbered,ruled,vlined]{algorithm2e}

\newtheorem{theorem}{Theorem}

\newtheorem{lemma}{Lemma}

\newtheorem{corollary}{Corollary}

\makeatletter
\def\ScaleIfNeeded{%
\ifdim\Gin@nat@width>\linewidth \linewidth \else \Gin@nat@width
\fi } \makeatother

\begin{document}
\title{Pinching-Antenna System (PASS) Enhanced Covert Communications}

\author{Hao~Jiang,~\IEEEmembership{Graduate Student Member,~IEEE,}  
Zhaolin~Wang,~\IEEEmembership{Member,~IEEE,} \\
and Yuanwei~Liu,~\IEEEmembership{Fellow,~IEEE}
\thanks{
H. Jiang, and Z. Wang are with the School of Electronic Engineering and Computer Science, Queen Mary University of London, London E1 4NS, U.K. (e-mail: \{hao.jiang; zhaolin.wang\}@qmul.ac.uk).

Y. Liu is with the Department of Electrical and Electronic Engineering, The University of Hong Kong, Hong Kong (e-mail: yuanwei@hku.hk).
}
}

\markboth{Manuscript}
{Manuscript}

\maketitle
\begin{abstract}
	A Pinching-Antenna SyStem (PASS)-assisted convert communication framework is proposed.
	PASS utilizes dielectric waveguides with freely positioned pinching antennas (PAs) to establish strong line-of-sight links.
	Capitalizing on this high reconfigurable flexibility of antennas, the potential of PASS for covert communications is investigated.
	1)~For the single-waveguide single-PA (SWSP) scenario, a closed-form optimal PA position that maximizes the covert rate is first derived.
	Subsequently, a one-dimensional power search is employed to enable low-complexity optimization for covert communications.
	With antenna mobility on a scale of meters, PASS can deal with the challenging situation of the eavesdropper enjoying better channel conditions than the legal user. 
	2)~For the multi-waveguide multi-PA (MWMP) scenario, the positions of multiple PAs are optimized to enable effective pinching beamforming, thereby enhancing the covert rate.
	To address the resultant multimodal joint transmit and pinching beamforming problem, a twin particle swarm optimization (TwinPSO) approach is proposed.
	Numerical results demonstrate that: i)~the proposed approaches can effectively resolve the optimization problems; ii)~PASS achieves a higher covert rate than conventional fixed-position antenna architectures; and iii)~with enhanced flexibility, the MWMP setup outperforms the SWSP counterpart. 	
\end{abstract}

\section{Introduction}
The last serval decades have witnessed a fast development of wireless communication technology, evolving from merely supporting basic functionalities, such as mobile calls and texting, to proliferating wide, versatile application scenarios, from live-streaming to virtual reality \cite{jiang2021road}.
As customers' daily requirements regarding transmission rates have been largely satisfied, the next-generation wireless communication technologies are expected to push the boundary of telecommunication to satisfy more advanced requirements.
Among these requirements, the most important one is security, as privacy-sensitive information is inevitability conveyed via ubiquitous wireless connectivity\cite{wu2018surveyPLS}.

The most straightforward way to secure transmission is via data cryptography techniques, which typically utilize passwords or public-private key authentication to protect data security at the upper layer of a wireless network. 
Although effective in most scenarios, data cryptography is difficult to apply to some emerging networking architectures, such as ad-hoc networks and wireless sensor networks, due to the key management issue and high-complexity encryption and deciphering process \cite{poor2017wireless, yang2015safeguarding}. 
Instead of focusing on upper layers, the physical-layer security (PLS) technique manipulates the characteristics of wireless channels to prevent eavesdroppers from decoding the protected transmissions.
Compared to cryptography techniques, PLS focuses on the lower physical layer, thus providing low complexity and scalable security mechanisms for wireless transmissions \cite{yang2015safeguarding}.
Recently, covert communication has emerged as a new security paradigm for security. 
Unlike data cryptography and conventional PLS techniques, covert communication, emerged as a novel physical-layer approach, conceals the transmission behavior of legal users.
In other words, legitimate transmission happens without being noticed by the malicious eavesdroppers.
Consequently, the level of security offered by covert communications will not be weakened by the superior deciphering ability of malicious users \cite{jin2025covert}.

To unleash the benefits of covert communications, lots of theoretical work has been dedicated to investigating covert communication capacity \cite{bash2013limits}, noise uncertainty modeling at the illegal user \cite{he2017on}, and performance limitation over different channel models \cite{wang2016fundmental}, which all confirm the superiority of covert communications.
On the engineering side, covert communication also spurs great enthusiasm throughout the community.
In particular, for hiding legitimate transmission, the random nature of the wireless environment is a perfect choice \cite{chen2023covert}, as it can disguise the perturbance caused by legitimate transmission. 
Following such an idea, the authors of \cite{hu2018covert} and \cite{shahzad2017covert} investigated covert communications by harnessing the large-scale and small-scale fading of the wireless channel, respectively.
Since environmental randomness induced by channels can only be manipulated with restricted flexibility, artificial noise is utilized at friendly jammers to realize more controllable transmission covertness.
This category is exemplified by \cite{soltani2018covert}, which demonstrates an increase in covert capacity from $\mathcal{O}(\sqrt{n})$ to $\mathcal{O} \left( \min \left\{ n,m^{\gamma /2}\sqrt{n} \right\} \right)$, where $n$, $\gamma$, and $m$ denote the number of channel uses, the path-loss exponent, and the density of jammers' distributions, respectively.
For long-range covert transmissions, unmanned aerial vehicles (UAVs) and relay technologies are widely adopted to mitigate the design dilemma between ``high transmit power needed for overcoming large path loss" and ``transmission covertness against eavesdroppers". 
For example, the authors of \cite{zhou2019joint} designed a UAV trajectory to ensure covertness under uncertainty in the eavesdropper's noise and location. Separately, \cite{wang2018covert} leveraged both relay assistance and channel uncertainty to protect private transmissions.

With the advent of the multiple-input and multiple-output (MIMO) era, the multi-antenna technology is widely exploited for security transmission \cite{chen2023covert}.
Intuitively, beamforming steers the majority of transmit power towards the legitimate user while minimizing leakage towards the eavesdropper.
In particular, the authors of \cite{zheng2019multi} investigated a covert communication network with the aid of centralized and distributed multi-antenna setups, demonstrating the effectiveness of multi-antenna technologies.
As a separate study, the authors of \cite{forouzesh2020covert} presented a beamforming optimization framework to deal with the uncertainty of the eavesdropper's location.
Furthermore, \cite{wang2021covert} dealt with the scenario when multiple antennas are equipped at all nodes in the classic covert communication setup.
As MIMO technology evolves, more and more antenna elements are being packed into a given area, driving a trend from MIMO toward gigantic MIMO \cite{emil2024enabling6gperformanceupper}.
However, such a trend is followed by increased hardware costs and design complexity, motivating the adoption of flexible antenna designs.
Different from conventional MIMO configurations, flexible antenna designs aim at reconfiguring the wireless channel itself, instead of passively adapting to it.
Within the scope of covert communications, antenna flexibility also offers significant advantages.
As the most exemplified technology, reconfiguration intelligent surfaces (RISs) aided covert communication has received considerable attention.
For example, the authors of \cite{wang2021intelligent} revealed the potential of RIS in enhancing the covertness of data transmission.
As further advancement, the authors of \cite{wang2025star} employed simultaneously transmitting and reflecting RIS (STAR-RIS) to achieve covert communications outdoors while simultaneously serving an indoor user.
Considering the near-field phenomenon, the authors of \cite{liu2024ris} exploited the beamfocusing effect to address the situation where the eavesdropper is located closer to the transmitter.
More recently, fluid antennas and movable antennas have emerged as new flexible antenna configurations, allowing the physical movement of antenna elements across the antenna panel.
As a pioneering work, the authors of \cite{liu2025movable} demonstrated that antenna mobility can enable a higher level of covertness compared to conventional fixed-position antenna designs, especially when the number of antennae is small.

Although these methods showcase the potential of reconfiguration on wireless channels, either by passive beamforming or antenna mobility, their flexibility remains limited.
Specifically, movable and fluid antenna technologies only allow antenna position adjustment merely at wavelength scale, thereby making them primarily suitable for mitigating small-scale fading.
For RIS, passive beamforming will inevitably cause the double-fading effect, which restricts the achievable performance gain through its implementation.
These limitations are calling for new, more flexible antenna technologies.
Recently, the pinching-antenna system (PASS) has become a promising candidate for unlocking the spatial flexibility of antenna elements.
In particular, PASS comprises single or multiple dielectric waveguides that carry the intended signals.
Attached to these waveguides, single or multiple separated dielectric elements, referred to as pinching antennas (PAs), radiate the signal from the waveguides into free space.
In practice, the first prototype of PASS was proposed and demonstrated by DOCOMO in \cite{suzuki2022pinching}, serving as the foundational hardware platform for subsequent theoretical research.
In contrast to fluid antenna, movable antenna, and RIS techniques, PASS mitigates large-scale fading by allowing part of the signal propagation to happen within low-attenuation waveguides, heralding a technical shift in wireless service from ``last mile'' to ``last meter'' \cite{liu2025pinching}.
Motivated by the advantages of PASS, numerous recent research efforts have been devoted to exploring its potential, such as performance analysis \cite{ouyang2025array,ding2024flexible, tyrovolas2025performance}, beamforming design \cite{ding2024flexible,wang2025modeling,guo2025gpassdeep}, beam training and channel estimation schemes \cite{lv2025beamtrainin, xiao2025channel}, and PLS \cite{sun2025physical}, among others.

In terms of covert communications, the benefits of applying PASS are two-fold: 
\emph{1) Enhanced Flexibility:} Unlike wavelength-scale mobility of fluid and movable antenna systems, the length of low-attenuation waveguides in PASS can span tens of meters, endowing PASS with an enhanced flexibility to mitigate large-scale fading in wireless channels. 
In terms of covert communications, when an unauthorized eavesdropper is located closer to the transmitter than the legal user, neither the conventional fixed-position nor movable antenna configurations can effectively conceal the transmission.
As a remedy, PASS addresses this challenge by emitting signals from the PA positioned right above the legitimate user. 
Furthermore, compared to RIS, the number of PAs can be arbitrarily modified at low costs, thus making PASS more scalable and flexible.
\emph{2) Low Implementation Cost:} Compared to relay- or UAV-aided covert communications, PASS can reduce the system complexity by eliminating the need for pre-defined relay protocols or high-complexity trajectory designs.
More importantly, unlike the sophisticated hardware required for relays or UAVs, only low-cost dielectric waveguides and PAs are needed for PASS, significantly reducing implementation costs.
To harness these benefits for covert communications, we investigate the joint PA position and beamforming optimization, for both the single-waveguide single-PA (SWSP) and multi-waveguide multi-PA (MWMP) scenarios.
The key contributions of this work are summarized as follows:
\begin{itemize}
	\item We propose a PASS-aided covert communication framework, where the transmitter, referred to as Alice, aims to deliver secret signals to a legal user, referred to as Bob, in the presence of a watchful eavesdropper, referred to as Willie.
	Besides, Willie's position and noise uncertainty are taken into account in the system model.

	\item We first investigate the SWSP case, where PA position and transmit power optimization are needed.
	To ensure covertness, we first introduce the concept of a forbidden zone to address Willie's position uncertainty issue.
	Building on this, the closed-form optimal solution of PA position that maximizes the covert rate is derived.
	Subsequently, a one-dimensional search algorithm is proposed to address the remaining transmit power optimization.
	By doing so, a low-complexity optimization framework for the SWSP scenario is developed.

	\item We then consider the MWMP case, which requires optimizations on multiple PA positions and beamforming design compared to the SWSP case.
	Due to the complexity of this scenario, the closed-form optimal solution to the antenna position is intractable.
	Instead, we employ a sampling method over Willie's possible positions to guarantee the covertness over the position uncertainty region.
	Finally, a twin particle swarm (TwinPSO) algorithm is proposed to optimize PA positions and beamforming jointly, which can further enhance the covert rate with higher flexibility compared to the SWSP scenario.

	\item We provide extensive simulation results to validate the superiority of PASS over several benchmarks.
	The results demonstrate the following points: 
	i)~The proposed approaches effectively leverage the flexibility of PASS to achieve a high covert rate.
	ii)~With the increased flexibility of the MWMP setup, PASS achieves a higher covert rate compared to the SWSP scenarios at the cost of higher optimization complexity.
	iii)~Versus conventional fixed-antenna MIMO, PASS offers significantly higher covert rates by enabling LoS propagation.
\end{itemize}

The remainder of the paper is organized as follows:
Section \ref{sect:swsp} introduces the SWSP setup and a one-dimensional search algorithm. Section \ref{sect:mwmp} addresses the more complex MWMP case and presents the TwinPSO algorithm. Numerical results are provided in Section \ref{sect:results}, and conclusions are drawn in Section \ref{sect:conlusion}.

\textit{Notations:}
Scalars, vectors, and matrices are denoted by the lower-case, bold-face lower-case, and bold-face upper-case letters, respectively.
$\mathbb{C}^{M \times N}$ and $\mathbb{R}^{M \times N}$ denote the space of $M \times N$ complex and real matrices, respectively.
$(\cdot)^\mathrm{T}$, $(\cdot)^*$, and $(\cdot)^\mathrm{H}$ denote the transpose, conjugate, and conjugate transpose, respectively.
$|\cdot|$ represents absolute value.
For a vector $\mathbf{a}$, $[\mathbf{a}]_i$ and $\left\| \mathbf{a} \right\| $ denote the $i$-th element and $2$-norm, respectively.
$\mathrm{j}=\sqrt{-1}$ denotes the imaginary unit.

\section{Using A Single Pinching Antenna on A Single Waveguide} \label{sect:swsp}
\begin{figure}[t!]
    \centering
    \includegraphics[width=0.9\linewidth]{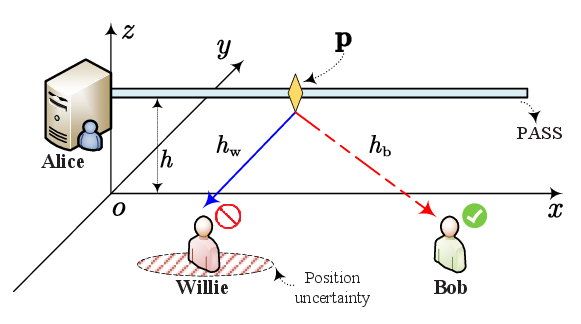}
    \caption{Illustration of covert communication system model in the SWSP case.}
    \label{fig:system_model_1}
\end{figure}
The covert communication system model is shown in Fig. \ref{fig:system_model_1}, where the transmitter (Alice) intends to transmit to a legitimate user (Bob) with the presence of a malicious detection node (Willie).
Alice has a single-waveguide PASS with a single PA to convey covert information to single-antenna Bob through downlink transmissions.
Alongside this legitimate transmission, single-antenna Willie aims to detect the covert transmission activity from Alice to Bob.
According to the classic covert communication model \cite{chen2023covert}, Alice needs to design the position of PA to hide secret transmission to Bob from Willie's detection.
To ensure transmission covertness, we consider the worst-case scenario.
More particularly, we assume that Willie knows the exact position of PA, whereas Alice knows Willie's rough position.
For the system topology, the coordinate origin is set at the feed point of Alice's waveguide and the waveguide of Alice is placed along the $x$-axis at the height of $h$.
The waveguide length at Alice is set to $L$, which confines the feasible positions of PA.
In addition, Willie and Bob are located on the $XOY$ plane.
A common narrowband is utilized for transmissions, and the channels are line-of-sight (LoS). 

\subsection{Channel Model and Signal Model}
The position of PA is specified by $\mathbf{p}=[x, 0, h] ^{\mathrm{T}}$, while the positions of Willie and Bob are expressed as $\mathbf{r}_{\mathrm{w}}=[ x_{\mathrm{w}},y_{\mathrm{w}},0] ^{\mathrm{T}}$ and $\mathbf{r}_{\mathrm{b}}=[x_{\mathrm{b}},y_{\mathrm{b}},0] ^{\mathrm{T}}$, respectively.
According to \cite{ouyang2025array} and \cite{ding2024flexible}, the channel coefficient from Alice to Bob and that from Alice to Willie can be respectively expressed as
\begin{align}
    h_{\mathrm{b}}&=\frac{\sqrt{\eta}e^{-\mathrm{j}k_{\mathrm{c}}\left\| \mathbf{r}_{\mathrm{b}}-\mathbf{p} \right\|}}{\left\| \mathbf{r}_{\mathrm{b}}-\mathbf{p} \right\|}, \\
    h_{\mathrm{w}}&=\frac{\sqrt{\eta}e^{-\mathrm{j}k_{\mathrm{c}}\left\| \mathbf{r}_{\mathrm{w}}-\mathbf{p} \right\|}}{\left\| \mathbf{r}_{\mathrm{w}}-\mathbf{p} \right\|},
\end{align}
where $k_{\mathrm{c}}\triangleq \frac{2\pi}{\lambda}$ denotes the free-space wavenumber with $\lambda$ being the free-space carrier wavelength, and $\eta =\frac{\lambda ^2}{16\pi ^2}$ denotes the propagation constant.
The phase shift of signal caused by the in-waveguide propagation is given by
\begin{align}
    w=e^{-\mathrm{j}k_{\mathrm{g}}\left( x-x_0 \right)},
\end{align}
where $x_0=0$ denotes the feed-point coordinate of the waveguide, $k_{\mathrm{g}}\triangleq \frac{2\pi}{\lambda _{\mathrm{g}}}$ denotes the wavenumber inside the waveguide, and $\lambda _{\mathrm{g}} = \lambda/n_{\mathrm{eff}}$ with $n_{\mathrm{eff}}$ being the effective refractive index of a dielectric waveguide \cite{pozar2012microwave}.
Jointly considering both free-space and in-waveguide propagations, the received downlink signal at Bob can be expressed as
\begin{align}
    y_{\mathrm{b}}&=\sqrt{P}h_{\mathrm{b}}w c_{\mathrm{b}}+n_{\mathrm{b}},
\end{align}
where $P$ denotes the transmit power of Alice, $c_{\mathrm{b}}$ denotes the normalized secret signal for Bob, and $n_{\mathrm{b}} \sim \mathcal{CN}(0, \sigma_{\mathrm{b}}^2)$ is the complex-valued additive white Gaussian noise with $\sigma_{\mathrm{b}}^2$ being the noise power.
Therefore, the signal-to-noise ratio (SNR) at Bob can be expressed as
\begin{align}
    \gamma _{\mathrm{b}}(x, P)=\frac{P\left| h_{\mathrm{b}}w \right|^2}{\sigma _{\mathrm{b}}^{2}}. \label{eq:snr}
\end{align}
According to \eqref{eq:snr}, when Bob's position is fixed, the throughput is determined by PA's $x$-coordinate and the transmit power.

Next, we model the signal received at Willie.
In light of the covert communication rationale \cite{chen2023covert,zhou2019joint}, Willie hypothesizes two situations, i.e., 1) $\mathcal{H}_0$: Alice is keeping silent, and 2) $\mathcal{H}_1$: Alice is transmitting to Bob.
Correspondingly, the received signal at Willie under the hypothesis is given by
\begin{align}
    y_{\mathrm{w}}=
\begin{cases}
	n_{\mathrm{w}}, & \mathcal{H}_0, \\
	\sqrt{P} h_{\mathrm{w}} w c_{\mathrm{b}} + n_{\mathrm{w}}, & \mathcal{H}_1,
\end{cases}
\end{align}
where $n_{\mathrm{w}} \sim \mathcal{CN}(0, \sigma_{\mathrm{w}}^2)$ denotes the complex-valued additive white Gaussian noise with power $\sigma_{\mathrm{w}}^2$.
\begin{figure*}[!t]
    \begin{align}\label{eq:xi}
   \xi =\begin{cases}
   	1,&		\mathrm{if}~\Gamma _{\mathrm{th}}<\bar{\sigma}_{\mathrm{w}}^{2}/\Delta _{\sigma}^{},\\
   	\ln \left( \Delta _{\sigma}^{}\bar{\sigma}_{\mathrm{w}}^{2}/\Gamma _{\mathrm{th}} \right) /2\ln \left( \Delta _{\sigma}^{} \right) ,&		\mathrm{if}~\bar{\sigma}_{\mathrm{w}}^{2}/\Delta _{\sigma}^{}\le \Gamma _{\mathrm{th}}<\bar{\sigma}_{\mathrm{w}}^{2}/\Delta _{\sigma}^{}+P\left| h_{\mathrm{w}}w \right|^2,\\
   	\ln \left( \Delta _{\sigma}^{2}\left( 1-P\left| h_{\mathrm{w}}w \right|^2/\Gamma _{\mathrm{th}} \right) \right) /2\ln \left( \Delta _{\sigma}^{} \right) ,&		\mathrm{if}~\bar{\sigma}_{\mathrm{w}}^{2}/\Delta _{\sigma}^{}+P\left| h_{\mathrm{w}}w \right|^2\le \Gamma _{\mathrm{th}}<\Delta _{\sigma}^{}\bar{\sigma}_{\mathrm{w}}^{2},\\
   	\ln \left( \Delta _{\sigma}^{}\left( \Gamma _{\mathrm{th}}-P\left| h_{\mathrm{w}}w \right|^2 \right) /\bar{\sigma}_{\mathrm{w}}^{2} \right) /2\ln \left( \Delta _{\sigma}^{} \right) ,&		\mathrm{if}~\Delta _{\sigma}^{}\bar{\sigma}_{\mathrm{w}}^{2}\le \Gamma _{\mathrm{th}}<\Delta _{\sigma}^{}\bar{\sigma}_{\mathrm{w}}^{2}+P\left| h_{\mathrm{w}}w \right|^2,\\
   	1,&		\mathrm{if}~\Gamma _{\mathrm{th}}\ge \Delta _{\sigma}^{}\bar{\sigma}_{\mathrm{w}}^{2}+P\left| h_{\mathrm{w}}w \right|^2.\\
   \end{cases}
\end{align}
    \hrulefill 
\end{figure*}

\subsection{Detection Performance Analysis}
Due to the uncooperative behavior of the malicious user, the transceivers cannot precisely know the distribution of additive noise and only know the noise power distribution.
Thus, we adopt the bounded noise uncertainty model in \cite{he2017on} and \cite{zhou2019joint}.
Specifically, the exact noise power $\sigma_{\mathrm{w}}^2$ follows a uniform distribution and is confined within a bounded range of $\sigma _{\mathrm{w},\mathrm{dB}}^{2}\in [\bar{\sigma}_{\mathrm{w},\mathrm{dB}}^{2}-\Delta _{\sigma ,\mathrm{dB}},\bar{\sigma}_{\mathrm{w},\mathrm{dB}}^{2}+\Delta _{\sigma ,\mathrm{dB}}]$, where $\bar{\sigma}_{\mathrm{w},\mathrm{dB}}^{2}$ denotes the nominal noise power and $\Delta _{\sigma ,\mathrm{dB}}$ controls the range of the uncertainty.
The subscript $\mathrm{dB}$ indicates that these quantities are expressed in decibels.
Therefore, we have $\sigma _{\mathrm{w},\mathrm{dB}}^{2}=10\log _{10}\left( \sigma _{\mathrm{w}}^{2} \right)$, $\bar{\sigma}_{\mathrm{w},\mathrm{dB}}^{2}=10\log _{10}\left( \bar{\sigma}_{\mathrm{w}}^{2} \right)$, and $\Delta _{\sigma ,\mathrm{dB}}=10\log _{10}\left( \Delta _{\sigma} \right)$ with $\Delta _{\sigma}>1$, respectively.
Given the uniform distribution of $\sigma _{\mathrm{w},\mathrm{dB}}^{2}$, the probability density function (PDF) of $\sigma _{\mathrm{w}}^{2}$ can be derived as
\begin{align}
    f_{\sigma _{\mathrm{w}}^{2}}\left( x \right) =\begin{cases}
	\frac{1}{2\ln \left( \Delta_{\sigma} \right) x}, &\mathrm{if}~\frac{1}{\Delta_{\sigma}}\bar{\sigma}_{\mathrm{w}}^{2}\le x \le \Delta_{\sigma} \bar{\sigma}_{\mathrm{w}}^{2},\\
	0. &\mathrm{otherwise}.\\
\end{cases}
\end{align}

To detect whether Alice is transmitting, the optimal test can be expressed as \cite{yan2018delay} 
\begin{align}
    \mathscr{T} \triangleq \frac{1}{N}\sum\nolimits_{n=1}^N{\left| y_{\mathrm{w}}\left[ n \right] \right|^2\underset{\mathcal{D} _0}{\overset{\mathcal{D} _1}{\gtrless}}\Gamma _{\mathrm{th}},} \label{eq:decision}
\end{align}
where $N$ denotes the total number of channel uses, and $\Gamma_{\mathrm{th}}$ is the preset detection threshold.
If $\mathscr{T}$ exceeds $\Gamma_{\mathrm{th}}$, Willie hypothesizes that Alice is transmitting signals to Bob, which is defined as $\mathcal{D}_1$; otherwise, Willie hypothesizes Alice keeps silent, which is defined as $\mathcal{D}_0$.
Letting $N \rightarrow +\infty$, the test $\mathscr{T}$ can be derived as 
\begin{align}
    \mathscr{T}=\begin{cases}
	\sigma _{\mathrm{w}}^{2}, &\mathcal{H} _0, \\
	P\left| h_{\mathrm{w}}w \right|^2+\sigma _{\mathrm{w}}^{2}, &\mathcal{H} _1. \label{eq:test}
\end{cases}
\end{align}
Based on \eqref{eq:decision} and \eqref{eq:test}, the detection error occurs at Willie includes two scenarios: 1) \emph{Miss Detection}: Alice is transmitting to Bob, while Willie wrongly believes Alice keeps silent; 2) \emph{False Alarm}: Alice keeps silent, while Willie wrongly infers Alice is transmitting.
The probabilities for these two scenarios can be respectively expressed as  
\begin{align}
	P_{\mathrm{M}}&=\mathbb{P} \left\{ \mathcal{D} _0\left| \mathcal{H} _1 \right. \right\} =\mathbb{P} \left\{ P\left| h_{\mathrm{w}}w \right|^2+\sigma _{\mathrm{w}}^{2}\le \Gamma _{\mathrm{th}} \right\}, \label{eq:miss_detection}\\
    P_{\mathrm{F}}&=\mathbb{P} \left\{ \mathcal{D} _1\left| \mathcal{H} _0 \right. \right\} =\mathbb{P} \left\{ \sigma _{\mathrm{w}}^{2}\ge \Gamma _{\mathrm{th}} \right\}. \label{eq:false_alarm}
\end{align}
Therefore, the total error rate at Willie, encapsulating miss detection and false alarming, can be computed by $\xi =P_{\mathrm{F}}+P_{\mathrm{M}}$.
From the perspective of a malicious user, the objective of Willie is obtaining the optimal detection threshold, denoted by $\Gamma _{\mathrm{w},\mathrm{th}}^{\mathrm{opt}}$, in the sense that the total error rate $\xi$ is minimal.
To this end, building on \eqref{eq:miss_detection} and \eqref{eq:false_alarm}, the total error rate $\xi$ can be expressed as \eqref{eq:xi} at the top of the next page.
By analysis the monotonicity of \eqref{eq:xi}, we have the following lemma:
\begin{lemma} \label{lemma:opt_detection}
    \normalfont
    \emph{(Optimal Detection Threshold at Willie)}    
    Given the exact position of PA, the optimal detection threshold at Willie, denoted by $\Gamma _{\mathrm{w, th}}^{\mathrm{opt}}$, and the corresponding minimal total error rate $\xi^{\rm{min}}_{\mathrm{w}}$ can be respectively given by
    \begin{align}
		\Gamma _{\mathrm{w},\mathrm{th}}^{\mathrm{opt}}&=\frac{\bar{\sigma}_{\mathrm{w}}^{2}}{\Delta _{\sigma}}+P\left| h_{\mathrm{w}}w \right|^2, \\
        \xi _{\mathrm{w}}^{\min}&=\frac{1}{2\ln (\Delta _{\sigma} )}\ln \left( \frac{\Delta _{\sigma}^{2}\bar{\sigma}_{\mathrm{w}}^{2}}{\bar{\sigma}_{\mathrm{w}}^{2}+P\left| h_{\mathrm{w}}w \right|^2\Delta _{\sigma}} \right). 
    \end{align}
\end{lemma}
\begin{IEEEproof}
    Please refer to \cite[Appendix A]{zhou2019joint}.
\end{IEEEproof}
According to \textbf{Lemma \ref{lemma:opt_detection}}, the optimal detection threshold and the total error rate can be achieved on Willie's side. 
However, on the transmitter side, Willie's position is typically hard to obtain due to its non-cooperative behavior.
In this case, Alice only knows Willie's rough position.
To model this case, we consider a bounded position error model, which can be expressed as
\begin{align}
    \mathbf{r}_{\mathrm{w}}\in \left[ \bar{\mathbf{r}}_{\mathrm{w}}-\Delta \mathbf{r}_{\mathrm{w}},\bar{\mathbf{r}}_{\mathrm{w}}+\Delta \mathbf{r}_{\mathrm{w}} \right] \triangleq \mathcal{G}, \label{eq:bounded_position_uncertainty}
\end{align}
where $\bar{\mathbf{r}}_{\rm w}$ is the coordinates of the center of the uncertainty region, and $\Delta \mathbf{r}_{\mathrm{w}}\triangleq \left[ \Delta x,\Delta y,0 \right] ^{\mathrm{T}}$ denotes the radius of this uncertainty region with $\left\| \Delta \mathbf{r}_{\mathrm{w}} \right\|= \Delta_{r}$.
Moreover, $\Delta_{r}$ denotes the radius of position uncertainty.
We assume that $\Delta_{r}$ is known at Alice.
Similar to Willie, to guarantee covertness, Alice must figure out the optimal detection threshold and the associated minimal total error rate. 
Nevertheless, due to the uncertainty of Willie's position, the minimal total detection rate at Willie's exact position cannot be obtained by Alice, thereby making it challenging to ensure transmission covertness.
To overcome this uncertainty, for each possible position of Willie $\mathbf{r}_{\rm w}$, Alice can compute its corresponding optimal detection threshold using \textbf{Lemma \ref{lemma:opt_detection}}.
By doing this repeatedly, the optimal detection threshold set for all possible ${\mathbf{r}}_{\mathrm{w}} \in \mathcal{G}$ can be obtained. 
Based on this set, the corresponding set containing all possible minimal total error rates can be computed and given by
\begin{align}
	\mathcal{E} \triangleq \left\{ \mathbf{r}_{\mathrm{w}}\in \mathcal{G} :\xi _{\mathrm{w}}^{\mathrm{min}} \right\}, \label{eq:minimal_error_rate_set}
\end{align}
Building on the above preliminaries, the covert communication problem and its solution will be presented in the upcoming subsection.

\subsection{Problem Formulation and Solution}
In covert communications, Alice aims to maximize the throughput at Bob and hide the legitimate transmissions from the detection of Willie by enlarging the total error rate at Willie.
However, due to the unknown precise coordinates of Willie, the true total error rate is contained in $\mathcal{E}$.
Thereby, the covert rate maximization problem can be formulated as
\begin{subequations} \label{problem_1}
    \begin{align}
        \max_{x, P} \quad &\log _2\left( 1+ \gamma _{\mathrm{b}}(x, P) \right) 
          \label{eq:p1-obj}\\
        \mathrm{s.t.} \quad & 0\le x \le L, \label{eq:p1-c1}\\
        & 0 \le P \le P_{\max},  \label{eq:p1-c2}\\
        & \xi \ge 1 - \rho_{\mathrm{w}}, \quad \forall \xi \in \mathcal{E}, \label{eq:p1-c3}
    \end{align}
\end{subequations}
where $P_{\max}$ denotes the total power budget at Alice.
The position constraint \eqref{eq:p1-c1} confines the feasible region of PA's position on the waveguide; the power constraint \eqref{eq:p1-c2} ensures that the transmit power cannot exceed the total power budget;
finally, the covertness constraint \eqref{eq:p1-c3} guarantees that the total error rate at Willie is larger than a predefined threshold $1-\rho_{\rm w}$ for any possible position.
It is noted that $\rho_{\rm w} \in [0, 1]$ is typically set to a small value to assure transmission covertness.

The crux of solving \eqref{problem_1} is dealing with the covertness constraint \eqref{eq:p1-c3} since it is enforced on a set rather than a specific value.
To address this constraint, we introduce the concept of forbidden zones, within which the transmission covertness will be violated.
On the flip side, by moving the PA outside the forbidden zone, the covertness can be satisfied, making \eqref{eq:p1-c3} be released.
In particular, the forbidden zone is specified by the following lemma:
\begin{lemma} \label{lemma:forbidden_zone}
    \normalfont
    \emph{(Forbidden Zone)} 
    Forbidden Zone is defined as the position region where the covertness constraint cannot always be satisfied within the uncertainty region of Willie's position.   
    Given that both $h<\sqrt{\frac{P\eta \Delta _{\sigma}}{\bar{\sigma}_{\mathrm{w}}^{2}\left( \Delta _{\sigma}^{2\rho _{\mathrm{w}}}-1 \right)}}$ and $\left| y_{\mathrm{w}} \right|<d_{\mathrm{tot}}$ hold, the forbidden zone can be specified by 
    \begin{align}
    	\mathcal{D} \triangleq \left[ x_{\mathrm{w}}-d_{\mathrm{tot}}\sin \left( \theta \right) , x_{\mathrm{w}}+d_{\mathrm{tot}}\sin \left( \theta \right) \right] = \left[ x_{\mathcal{D}}^{-},x_{\mathcal{D}}^{+} \right].
    \end{align}
	where 
	\begin{align}
		d_{\mathrm{tot}}&\triangleq \sqrt{\frac{P\eta \Delta _{\sigma}}{\bar{\sigma}_{\mathrm{w}}^{2}\left( \Delta _{\sigma}^{2\rho _{\mathrm{w}}}-1 \right)}-h}+\Delta _r, \\
		\theta &=\mathrm{arc}\cos \left\{ \frac{\left| y_{\mathrm{w}} \right|}{d_{\mathrm{tot}}} \right\}.
	\end{align}
	Otherwise, there is no forbidden zone for PA's position.
\end{lemma}
\begin{IEEEproof}
    Please refer to Appendix \ref{forbidden_zone_proof}.
\end{IEEEproof}
With \textbf{Lemma \ref{lemma:forbidden_zone}}, the original constraint \eqref{eq:p1-c3} can be converted as an equivalent constraint on PA's position.
Thereby, the original problem can be rewritten as
\begin{subequations} \label{problem_1a}
	\begin{align}
		\max_{x,P} \quad &\log _2\left( 1+\gamma _{\mathrm{b}}(x,P) \right) \\
		\mathrm{s.t.} \quad & \eqref{eq:p1-c1}~\text{and}~\eqref{eq:p1-c2}, \notag\\
		& x \notin \mathcal{D}.	\label{eq:p1a-c3}
	\end{align}
\end{subequations}
Due to the LoS transmission condition, for given transmit power $P$, the highest throughput can be achieved when PA is closest to the position of Bob.
Therefore, when transmit power $P$ is fixed, the sub-problem for PA position optimization can be formulated by 
\begin{subequations} \label{sub_problem_1}
	\begin{align}
		\min_{x} \quad & \left\| \mathbf{r}_{\mathrm{b}}-\mathbf{p} \right\| \label{sub-problem:x}\\
		\mathrm{s.t.} \quad & \eqref{eq:p1-c1}~\text{and}~\eqref{eq:p1a-c3}, \notag
	\end{align}
\end{subequations}
The optimal solution to \eqref{sub_problem_1} can be found at the feasible position closest to Bob.
For discussion simplicity, we combine constraints \eqref{eq:p1-c1} and \eqref{eq:p1a-c3} as the following constraint:
\begin{align}
	x\in \mathcal{D} \cap \mathcal{S} \triangleq \mathcal{F},
\end{align}
where $\mathcal{S}\triangleq [0, L]$ denotes the coordinate range of the waveguide.
Therefore, the optimal solution for \eqref{sub_problem_1} is given by the following theorem:
\begin{theorem}\label{theorem:optimal solution}
	\normalfont
	\emph{(Optimal PA's Position)} 
	For a given transmit power $P$, the optimal solution to sub-problem \eqref{sub_problem_1} can be specified by
	\begin{align}
	x^{\mathrm{opt}} = 
	\begin{cases}
		\mathrm{\emptyset}, & \text{if } \mathcal{S} \subseteq \mathcal{D}, \\
		x_{\mathrm{b}}, & \text{if } \mathcal{S} \nsubseteq \mathcal{D} \text{ and } x_{\mathrm{b}} \in \mathcal{F}, \\
		\mathop{\mathrm{arg}\min}\limits_{x \in \mathcal{A}} \left| x - x_{\mathrm{b}} \right|, & \text{if } \mathcal{S} \nsubseteq \mathcal{D} \text{ and } x_{\mathrm{b}} \notin \mathcal{F}.
	\end{cases}
  	\end{align}
	where $\mathcal{A}$ denotes the set containing all the boundary points of feasible set $\mathcal{F}$.
\end{theorem}
\begin{IEEEproof}
	Due to the LoS propagation, the throughput at Bob depends on the distance between PA and Bob.
	When $\mathcal{S} \subseteq \mathcal{D}$, the feasible region of PA's position is entirely included within the forbidden zone, implying that \eqref{sub_problem_1} has no feasible solution.
	Conversely, when $\mathcal{S} \nsubseteq \mathcal{D}$, there would be a feasible set denoted by $\mathcal{F}$ existing.
	If $x_{\rm b} \in \mathcal{F}$, the minimum distance between the PA and Bob is achieved at $x=x_{\rm b}$.
	Otherwise, if $x_{\rm b} \notin \mathcal{F}$, the minimum distance can be attained at the boundary of $\mathcal{F}$ that is closest to Bob.
	Moreover, jointly considering the above three conditions, the optimal solution is presented in \textbf{Theorem \ref{theorem:optimal solution}}, which completes this proof. 
\end{IEEEproof}
With \textbf{Theorem \ref{theorem:optimal solution}}, we have the closed-form optimal PA position for a given transmit power $P$.
It is important to note that the boundary of the forbidden zone is determined by the transmit power.
Consequently, when there is no feasible solution, one can continuously tune down the transmit power until a feasible solution is obtainable.
Moreover, the objective function \eqref{eq:p1-obj} is solely determined by transmit power $P$.
Therefore, in order to obtain the global optimal solution, we present a one-dimensional (1D) linear search algorithm.
Besides, it is apparent that the boundary of the forbidden region is a monotonically increasing function with $P$.
Therefore, in order to reduce complexity, this 1D search began with a small transmit power.
As $P$ is tuned larger, the condition $\mathcal{S} \subseteq \mathcal{D}$ will be met, indicating no feasible solution.
At this point, continuously enlarging $P$ is ineffective, implying the 1D search should be terminated once $\mathcal{S} \subseteq \mathcal{D}$.
Hence, based on this analysis, the overall 1D-search optimization algorithm is summarized by \textbf{Algorithm \ref{alg:swsp_power_searching}}.
\begin{algorithm}[t!]
	\SetAlgoLined
	\small
	\caption{1D Linear Search Algorithm for SWSP-aided Covert Communication}
	\label{alg:swsp_power_searching}
	\KwIn{
		Precise position of Bob $\mathbf{r}_{\mathrm{b}}$; 
		Approximated position of Willie given by $(\bar{\mathbf{r}}_{\rm w}, \Delta_r)$; 
		Total power budget $P_{\max}$;
		Noise power uncertainty given by $(\bar{\sigma}_{\rm w}^2, \Delta_{\sigma})$; 
		Power search resolution $\Delta_{P}$; 
		Covertness requirement parameter $\rho_{\rm w}$.
	}
	\KwOut{
		Optimized transmit power $P^{\rm opt}$, 
		Optimized PA position $\mathbf{p}^{\rm opt}$.
	}
	\DontPrintSemicolon
	\SetKwFunction{FReturn}{Return}
	
	\textbf{Initialization}: 
	Set iteration index $t \leftarrow 1$, 
	initialize $P^{(0)} \leftarrow 0$, 
	$\mathbf{p}^{(0)} \leftarrow [0, 0, h]^{\mathrm{T}}$, 
	and $R^{\rm opt} \leftarrow 0$. 
	
	\For{$P^{(t)} \le P_{\max}$}{
		Compute the optimal $x$-coordinate of the PA, $x^{\mathrm{opt}}$, 
		according to \textbf{Theorem \ref{theorem:optimal solution}}.\;
		
		\eIf{$x^{\mathrm{opt}}$ is not $\mathrm{\emptyset}$}{
			Construct the 3D position vector of the PA via $\mathbf{p}^{(t)} \leftarrow \bigl[x^{\mathrm{opt}},\, 0,\, h\bigr]^{\mathrm{T}}$\;
			
			Compute the covert rate $R^{(t)}$ via \eqref{eq:p1-obj}.\;
		}{
			\Return{$P^{\rm opt}$ and  $\mathbf{p}^{\rm opt}$}\;
		}
		\If{$R^{(t)} \ge R^{\rm opt}$}{
			Update $P^{\rm opt} \leftarrow P^{(t)}$, 
			$\mathbf{p}^{\rm opt} \leftarrow \mathbf{p}^{(t)}$, 
			and $R^{\rm opt} \leftarrow R^{(t)}$\;
		}
		Increase transmit power $P^{(t+1)}\leftarrow P^{(t)} + \Delta_{P}$\;
		Increment index $t \leftarrow t + 1$; 
	}
	\Return{$P^{\rm opt}$ and $\mathbf{p}^{\rm opt}$}
\end{algorithm}

\subsection{Optimality and Complexity Analysis}  
\subsubsection{Optimality} In this joint position and transmit power optimization problem, the closed-form solution for the position sub-problem is derived, thus ensuring its optimality.
Hence, the optimality of the joint problem depends on optimality of the transmit power.
Due to the exhaustive search algorithm, the optimality depends on the search resolution of transmit power, i.e., $\Delta_{P}$.
Specifically, only a local optimal point can be found when the resolution is coarse.
Otherwise, the global optimal solution can be guaranteed when the resolution is high.

\subsubsection{Complexity} Due to the closed-form optimal PA position, the computational complexity for this step is $\mathcal{O}(1)$.
For the transmit power optimization process, the computational complexity of the 1D linear search can be quantified as $\mathcal{O}(\frac{1}{\Delta_P })$.
Therefore, the overall complexity is computed by $\mathcal{O}(\frac{1}{\Delta_P })$.

\section{Using Multiple Pinching Antennas on Multiple Waveguides} \label{sect:mwmp}
\begin{figure}[t!]
	\centering
	\includegraphics[width=1\linewidth]{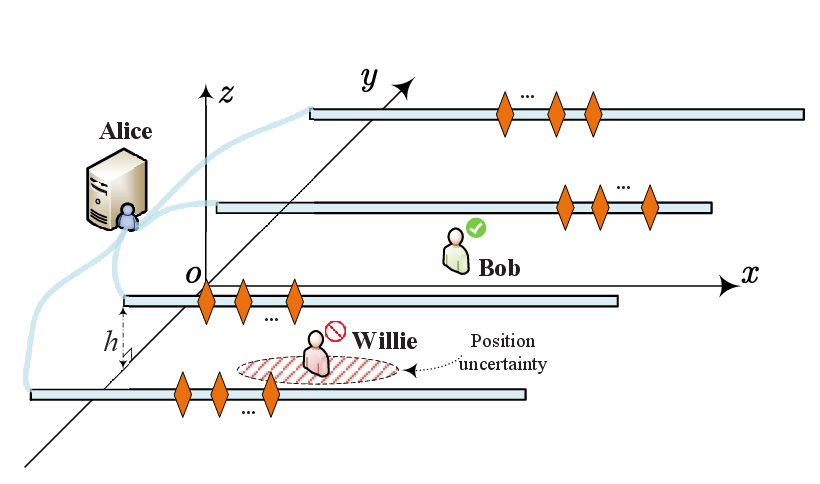}
	\caption{Illustration of covert communication system model in the MWMP case.}
	\label{fig:system_model_2}
\end{figure}
In this section, a MWMP PASS-aided covert communication system is considered.
The system model is illustrated in Fig. \ref{fig:system_model_2}, where PASS contains $N$ dielectric waveguides with each fed by a dedicated RF chain and equipped with $M$ PAs.
For the layout of PASS, the waveguides are placed along the $x$-axis at the height of $h$, and the coordinate origin is set at the center point of the line that connects all the feed points of the waveguides. 
Therefore, the position of the $m$-th PA on the $n$-th waveguide is given by $\mathbf{p}_{n,m}=[x_{n,m}, y_{n}, h]$, where $n\in\mathcal{M}\triangleq\{1,2,...,N\}$ and $m\in\mathcal{M}\triangleq\{1,2,...,M\}$.
Moreover, the legitimate user Bob and the malicious user Willie are located on the $XOY$ plane, and their coordinates are specified by $\mathbf{r}_{\mathrm{b}}=[x_{\mathrm{b}},y_{\mathrm{b}},0] ^{\mathrm{T}}$ and $\mathbf{r}_{\mathrm{w}}=[ x_{\mathrm{w}},y_{\mathrm{w}},0] ^{\mathrm{T}}$, respectively.
In contrast to the SWSP case, beamforming design must be considered alongside the PA position optimization due to the adoption of multiple-antenna technology.

\subsection{Channel Model and Signal Model}
Letting $c_{\mathrm{b}} \in \mathbb{C}$ be the intended signal for Bob with unit power, the transmitted signal at Alice can be expressed as
\begin{align}
	\mathbf{s}=\mathbf{w}c_{\mathrm{b}}\in \mathbb{C} ^{N\times 1},
\end{align}
where $\mathbf{w}\in \mathbb{C} ^{N\times 1}$ denotes the beamforming vector.
Let $\mathbf{x}_n\triangleq \left[ x_{n,1},x_{n,2},...,x_{n,M} \right] ^{\mathrm{T}}\in \mathbb{R} ^{M\times 1}$ be the $x$-axis coordinate vector for all PAs on the $n$-th waveguide.
The in-waveguide channel vector, representing the phase shifts induced by propagation within waveguides, is given by
\begin{align}
	\mathbf{g}\left( \mathbf{x}_n \right) =\left[ \alpha _{n,1}e^{-\mathrm{j}k_{\mathrm{g}}x_{n,1}},...,\alpha _{n,M}e^{-\mathrm{j}k_{\mathrm{g}}x_{n,M}} \right] ^{\mathrm{T}}\in \mathbb{C} ^{M\times 1},
\end{align}
where $\alpha _{n,m}$ denotes the power coefficient for the $m$-th PA on the $n$-th waveguide.
Here, we utilize the equal power model in \cite{wang2025modeling}, where PAs on the same waveguide share equal power, i.e., $\alpha _{n,m}=1 / \sqrt{M}$ for $\forall~m, n$.
In addition to the in-waveguide propagation, the free-space propagation also needs modeling.
Considering the positions of PAs and the LoS channel condition, the free-space channel vector for Bob and Willie concerning the $n$-th waveguide can be respectively given by 
\begin{align}
	\mathbf{h}_{\mathrm{b}}\left( \mathbf{x}_n \right) &=\left[ \frac{\sqrt{\eta}e^{-\mathrm{j}k_cr_{\mathrm{b},n,1}}}{r_{\mathrm{b},n,1}},...,\frac{\sqrt{\eta}e^{-\mathrm{j}k_cr_{\mathrm{b},n,M}}}{r_{\mathrm{b},n,M}} \right] ^{\mathrm{T}}\in \mathbb{C} ^{M\times 1},
	\\
	\mathbf{h}_{\mathrm{w}}\left( \mathbf{x}_n \right) &=\left[ \frac{\sqrt{\eta}e^{-\mathrm{j}k_cr_{\mathrm{w},n,1}}}{r_{\mathrm{w},n,1}},...,\frac{\sqrt{\eta}e^{-\mathrm{j}k_cr_{\mathrm{w},n,M}}}{r_{\mathrm{w},n,M}} \right] ^{\mathrm{T}}\in \mathbb{C} ^{M\times 1},
\end{align} 
where $r_{\mathrm{b},n,m}\triangleq \sqrt{\left( x_{\mathrm{b}}-x_{n,m} \right) ^2+\left( y_{\mathrm{b}}-y_n \right) ^2+h^2}$ and $r_{\mathrm{w},n,m}\triangleq \sqrt{\left( x_{\mathrm{w}}-x_{n,m} \right) ^2+\left( y_{\mathrm{w}}-y_n \right) ^2+h^2}$.
Then, jointly considering the in-waveguide and free-space propagations for all waveguides, the received signal at Bob can be compactly formulated as
\begin{align}
	y_{\mathrm{b}}&=\mathbf{h}_{\mathrm{b}}^{\mathrm{H}}\left( \mathbf{X} \right) \mathbf{G}\left( \mathbf{X} \right) \mathbf{w}c_{\mathrm{b}}+n_{\mathrm{b}},
\end{align}
where $n_{\mathrm{b}}\sim {\mathcal{CN}(0, \sigma_{\rm b}^2)}$ denotes the complex-valued additive white Gaussian noise, $\mathbf{X}\triangleq \left[ \mathbf{x}_1,\mathbf{x}_2,...,\mathbf{x}_N \right] \in \mathbb{R} ^{M\times N}$ is the $x$-coordinate matrix of all PAs, $\mathbf{h}_{\mathrm{b}}^{}\left( \mathbf{X} \right) \triangleq \left[ \mathbf{h}_{\mathrm{b}}^{\mathrm{T}}\left( \mathbf{x}_1 \right) ,\mathbf{h}_{\mathrm{b}}^{\mathrm{T}}\left( \mathbf{x}_2 \right) ,...,\mathbf{h}_{\mathrm{b}}^{\mathrm{T}}\left( \mathbf{x}_N \right) \right] ^{\mathrm{T}}\in \mathbb{C} ^{NM\times 1}
$ denotes the overall free-space channel vector of Bob, and $\mathbf{G}\left( \mathbf{X} \right)$ represents the overall in-waveguide channel matrix, specified by
\begin{align}
	\mathbf{G}\left( \mathbf{X} \right) \triangleq \left[ \begin{matrix}
		\mathbf{g}\left( \mathbf{x}_1 \right)&		\mathbf{0}&		\mathbf{0}&		\mathbf{0}\\
		\mathbf{0}&		\mathbf{g}\left( \mathbf{x}_2 \right)&		\cdots&		\mathbf{0}\\
		\vdots&		\vdots&		\ddots&		\mathbf{0}\\
		\mathbf{0}&		\mathbf{0}&		\mathbf{0}&		\mathbf{g}\left( \mathbf{x}_N \right)\\
	\end{matrix} \right] \in \mathbb{C} ^{NM\times N}.
\end{align}
Therefore, the received SNR at Bob can be expressed as 
\begin{align}
	\gamma _{\mathrm{b}}(\mathbf{X}, \mathbf{w})=\frac{\left| \mathbf{h}_{\mathrm{b}}^{\mathrm{H}}\left( \mathbf{X} \right) \mathbf{G}\left( \mathbf{X} \right) \mathbf{w} \right|^2}{\sigma _{\mathrm{b}}^{2}}.
\end{align}
Then, there are two hypotheses at Willie, corresponding to whether Alice is transmitting messages to Bob or not, which can be respectively given by
\begin{align}
	y_{\mathrm{w}}=\begin{cases}
		n_{\mathrm{w}}, &\mathcal{H} _0, \\
		\mathbf{h}_{\mathrm{w}}^{\mathrm{H}}\left( \mathbf{X} \right) \mathbf{G}\left( \mathbf{X} \right) \mathbf{w}c_{\mathrm{b}}+n_{\mathrm{w}},&\mathcal{H} _1,\\
	\end{cases}
\end{align}
where $\mathcal{H} _0$ represents that Alice keeps silence, while $\mathcal{H} _1$ indicates that Alice is transmitting.

\subsection{Detection Performance Analysis}
Similar to the SWSP case, we assume that the Alice only knows the bounded region of the noise power at Willie, which is specified by $\sigma _{\mathrm{w},\mathrm{dB}}^{2}\in [ \bar{\sigma}_{\mathrm{w},\mathrm{dB}}^{2}-\Delta_{\sigma, \mathrm{dB}},\bar{\sigma}_{\mathrm{w},\mathrm{dB}}^{2}+\Delta_{\sigma, \mathrm{dB}} ]$ with $\bar{\sigma}_{\mathrm{w},\mathrm{dB}}^{2}$ and $\Delta_{\sigma, \mathrm{dB}}$ being the nominal noise power and uncertainty radius, respectively.
Building on the above, the test $\mathscr{T}$ can be expressed as
\begin{align}
	\mathscr{T}=\begin{cases}
		\sigma _{\mathrm{w}}^{2},&		\mathcal{H} _0,\\
		\left| \mathbf{h}_{\mathrm{w}}^{\mathrm{H}}\left( \mathbf{X} \right) \mathbf{G}\left( \mathbf{X} \right) \mathbf{w} \right|^2+\sigma _{\mathrm{w}}^{2},&		\mathcal{H} _1.\\
	\end{cases} \label{eq:test-mwmp}
\end{align}
Based on \eqref{eq:test-mwmp}, the miss detection and false alarm probabilities for the MWMP case can be expressed as
\begin{align}
	P_{\mathrm{M}}&=\mathbb{P} \left\{ \mathcal{D} _0\left| \mathcal{H} _1 \right. \right\} =\mathbb{P} \left\{ \left| \mathbf{h}_{\mathrm{w}}^{\mathrm{H}}\left( \mathbf{X} \right) \mathbf{G}\left( \mathbf{X} \right) \mathbf{w} \right|^2+\sigma _{\mathrm{w}}^{2}\le \Gamma _{\mathrm{th}} \right\},
	\\
	P_{\mathrm{F}}&=\mathbb{P} \left\{ \mathcal{D} _1\left| \mathcal{H} _0 \right. \right\} =\mathbb{P} \left\{ \sigma _{\mathrm{w}}^{2}\ge \Gamma _{\mathrm{th}} \right\}.
\end{align}
Thus, the total error rate at Willie, consisting of both miss detection and false alarm, can be expressed as $\xi =P_{\mathrm{F}}+P_{\mathrm{M}}$.

Similar to the SWSP case, we assume that Willie knows the exact position of the PAs, which represents the worst-case scenario.
Hence, from Willie's perspective, the following lemma is proposed to obtain the optimal detection threshold and the associated minimal total error rate.
 \begin{lemma} \label{lemma:opt_detection_mwmp}
 	\normalfont
 	\emph{(Optimal Detection Threshold at Willie)}    
 	Given the exact position of PAs, the optimal detection threshold at Willie, denoted by $\Gamma _{\mathrm{w, th}}^{\mathrm{opt}}$, and the corresponding minimal total error rate $\xi^{\rm{min}}_{\mathrm{w}}$ can be respectively given by
 	\begin{align}
 		\Gamma _{\mathrm{w},\mathrm{th}}^{\mathrm{opt}}&=\frac{\bar{\sigma}_{\mathrm{w}}^{2}}{\Delta_{\sigma}}+\left| \mathbf{h}_{\mathrm{w}}^{\mathrm{H}}\left( \mathbf{X} \right) \mathbf{G}\left( \mathbf{X} \right) \mathbf{w} \right|^2 \\
 		\xi _{\mathrm{w}}^{\min}&=\frac{1}{2\ln (\Delta _{\sigma} )}\ln \left( \frac{\Delta _{\sigma}^{2}\bar{\sigma}_{\mathrm{w}}^{2}}{\bar{\sigma}_{\mathrm{w}}^{2}+\Delta _{\sigma}\left| \mathbf{h}_{\mathrm{w}}^{\mathrm{H}}\left( \mathbf{X} \right) \mathbf{G}\left( \mathbf{X} \right) \mathbf{w} \right|^2} \right). \label{eq:total_error_rate_ma}
 	\end{align}
 \end{lemma}
 \begin{IEEEproof}
 	This proof is the same as that in \textbf{Lemma \ref{lemma:opt_detection}}, and thus omitted here for brevity.
 \end{IEEEproof}
Unlike Willie, Alice only knows Willie's approximate position.
Therefore, the bounded position error can be specified by \eqref{eq:bounded_position_uncertainty}.
Correspondingly, accounting for the position uncertainty region $\mathcal{G}$ defined by \eqref{eq:minimal_error_rate_set}, the set containing all possible minimal total error rates, denoted by $\mathcal{E}$, can be computed by applying the same method used in \eqref{eq:minimal_error_rate_set} across all the possible Willie's locations in $\mathcal{G}$.

\subsection{Problem Formulation and Solution}
Based on the above preliminaries, the optimization problem is formulated in this subsection.
Similar to the SWSP scenario, Alice's objective is to convey as much information to Bob as possible without being detected by malicious Willie.
Therefore, the covert rate maximization problem is formulated as 
\begin{subequations} \label{problem_2}
	\begin{align}
		\max_{\mathbf{X},\mathbf{w}} \quad &\log _2\left( 1+\gamma _{\mathrm{b}}\left( \mathbf{X},\mathbf{w} \right) \right) 
		\label{eq:p2-obj}\\
		\mathrm{s}.\mathrm{t}.\quad &0\le x_{n,m}\le L,\forall m\in \mathcal{M} ,n\in \mathcal{N} \label{eq:p2-c1}\\
		&\Delta _{n,m}\triangleq x_{n,m}-x_{n,m-1}\ge \Delta _{\min},\forall m\in \mathcal{M}, \label{eq:p2-c2} \\
		& \left\| \mathbf{w} \right\|^{2}\le P_{\max},  \label{eq:p2-c3}\\
		& \xi \ge 1 - \rho_{\mathrm{w}}, \quad \forall \xi \in \mathcal{E}\label{eq:p2-c4}
	\end{align}
\end{subequations}
where the first position constraint \eqref{eq:p2-c1} regulates the feasible region of PA's position within the waveguide; the second position constraint \eqref{eq:p2-c2} enforces a minimum separation between PAs, denoted by $ \Delta _{\min}$, to prevent mutual coupling; the power constraint \eqref{eq:p2-c3} ensures the designed beamforming vector satisfying the total power budget; and the covertness constraint \eqref{eq:p2-c4} guarantees that the total error rate at Willie exceeds a predefined threshold.
It is noted that, compared to the SWSP scenario, \eqref{problem_2} introduces beamforming design on top of position optimization, which can enhance the flexibility of PASS.
To make this problem more tractable, we first simplify the constraints.
Given the expression of total error rate in \eqref{eq:total_error_rate_ma}, constraint \eqref{eq:p2-c4} can be converted into an equivalent constraint on beam gain at Willie, which is specified by
\begin{align}
	&\frac{1}{2\ln (\Delta _{\sigma} )}\ln \left( \frac{\Delta _{\sigma}^{2}\bar{\sigma}_{\mathrm{w}}^{2}}{\bar{\sigma}_{\mathrm{w}}^{2}+\Delta _{\sigma}\left| \mathbf{h}_{\mathrm{w}}^{\mathrm{H}}\left( \mathbf{X} \right) \mathbf{G}\left( \mathbf{X} \right) \mathbf{w} \right|^2} \right) \ge 1-\rho _{\mathrm{w}},\notag \\
	&g_{\mathrm{w}}\left( \mathbf{X},\mathbf{w} \right) \triangleq \left| \mathbf{h}_{\mathrm{w}}^{\mathrm{H}}\left( \mathbf{X} \right) \mathbf{G}\left( \mathbf{X} \right) \mathbf{w} \right|^2\le \frac{\bar{\sigma}_{\mathrm{w}}^{2}}{\Delta _{\sigma}}\left( \Delta _{\sigma}^{2\rho _{\mathrm{w}}}-1 \right) =\Gamma _{\mathrm{w}}, \label{eq:beam_gain}
\end{align}
where $g_{\mathrm{w}}\left( \mathbf{X},\mathbf{w} \right)$ represents the beam gain at Willie and $\Gamma _{\mathrm{w}}$ denotes the allowed maximal beam gain at Willie.
Recall that the channel vector $\mathbf{h}_{\mathrm{w}}$ is also parameterized by Willie's exact position $\mathbf{r}_{\mathrm{w}}$, which is unknown to Alice.
Therefore, to ensure \eqref{eq:beam_gain} holds for all the possible positions in $\mathcal{G}$, we reformulate \eqref{eq:p2-c4} as $\max _{\mathbf{r}_{\mathrm{w}}\in \mathcal{G}}g_{\mathrm{w}}\left( \mathbf{X},\mathbf{w} \right) \le \Gamma _{\mathrm{w}}$.
To decouple the optimization of $P$ and the phase of $\mathbf{w}$, we decompose $\mathbf{w}$ as $\mathbf{w}=\sqrt{P}\tilde{\mathbf{w}}$, where $\tilde{\mathbf{w}}$ is the normalized beamforming vector satisfying $\left\| \tilde{\mathbf{w}} \right\|^{2}=1$.
Consequently, we have $g_{\mathrm{w}}\left( \mathbf{X},\mathbf{w} \right) =Pg_{\mathrm{w}}\left( \mathbf{X},\tilde{\mathbf{w}} \right)$.
To optimize $P$, we present the following theorem to convert the inequality constraint into an equality counterpart.
 
 \begin{theorem}\label{theorem:optimal_transmit_power}
 	\normalfont
 	\emph{(Optimal Transmit Power)} Given that the phases of the beamforming vector are fixed, the optimal transmit power can be expressed as 
 	\begin{align}
		P^{\mathrm{opt}}=\min \left\{ \frac{\Gamma _{\mathrm{w}}}{g_{\mathrm{w}}^{\max}\left( \mathbf{X},\tilde{\mathbf{w}} \right)},P_{\max} \right\},
 	\end{align}
	where $g_{\mathrm{w}}^{\max}\left( \mathbf{X},\tilde{\mathbf{w}} \right) \triangleq \mathop {\max }_{\mathbf{r}_{\mathrm{w}}\in \mathcal{G}}~g_{\mathrm{w}}^{}\left( \mathbf{X},\tilde{\mathbf{w}} \right) $.
 \end{theorem}
 \begin{IEEEproof}
 	Please refer to Appendix \ref{optimal_transmit_power}.
 \end{IEEEproof}
 Unlike the SWSP case, it is challenging, if not impossible, to compute the forbidden zone where the constraint \eqref{eq:p2-c4} does not hold.
 To address this problem, we propose a sampling method.
 In practice, due to the wide beams emitted by PASS, the beam gains across Willie's position uncertainty region tend to be similar. 
 Therefore, by sampling points within the position uncertainty region, we can ensure that the constraint \eqref{eq:p2-c4} holds for most of Willie's possible positions.
 Motivated by this, we propose the following sampling strategy:
 \begin{align}
	\mathcal{C} \triangleq \bigg\{ \left. k \in \left\{ 1, \ldots, K \right\} \, \right|\, 
	& \mathbf{r}_{\mathrm{w}} \pm \left[ k \Delta_r / K, 0, h \right]^{\mathrm{T}}, \nonumber \\
	&\qquad \mathbf{r}_{\mathrm{w}} \pm \left[ 0, k \Delta_r / K, h \right]^{\mathrm{T}} \bigg\}, \label{eq:sampling}
\end{align}
 where $K$ is the total number of grids along one direction. 
 The sampling strategy in \eqref{eq:sampling} can be visualized as a set of evenly spaced concentric rings centered at $\bar{\mathbf{r}}_{\mathrm{w}}$.
 By applying this sampling method, constraint \eqref{eq:p2-c4} can be further converted to $g_{\mathrm{w}}^{\max}\left( \mathbf{X},\tilde{\mathbf{w}} \right) \simeq \mathop {\max }_{\mathbf{r}_{\mathrm{w}}\in \mathcal{C}}~g_{\mathrm{w}}^{}\left( \mathbf{X},\tilde{\mathbf{w}} \right) $.

 For PA position optimization, we fix the spacing between PAs on the same waveguide to $\Delta_x$, which satisfies $\Delta_x \ge \Delta_{\min}$.
 Thereby, instead of optimizing all PA positions, i.e., $\mathbf{X}$, we only need to optimize the initial position of PAs on each waveguide.
 Let the initial PA position vector be $\mathbf{x}_{\mathrm{init}}\triangleq \left[ x_{1,1},x_{2,1},...,x_{N,1} \right] ^{\mathrm{T}}\in \mathbb{R} ^{N\times 1}$.
 Considering constraints \eqref{eq:p2-c1} and \eqref{eq:p2-c2}, the updated position constraint on $\mathbf{x}_{\mathrm{init}}$ can be expressed as 
 \begin{align}
 	0\le x_{n,1}\le L-\Delta _x\left( M-1 \right) ,~~\forall n\in \mathcal{N}.  \notag
 \end{align}
The reasons for regulating the PA layout are two-fold: i)~\emph{Reduction in Computational Complexity:} Existing PASS studies that optimize all PA positions inevitably incur high computational complexity due to element-wise position optimization.
In fact, once the mutual coupling effect is overcome by setting $\Delta_{n,m}\ge \Delta_{\min}$, the performance gain obtained by optimizing the interspaces between PAs can be limited as verified by \cite{guo2025gpassdeep}.
To balance performance and complexity, we focus on optimizing the positions of the first PAs across waveguides.
 ii)~\emph{Compatibility with Realistic Hardware Model for PASS:} Prior work on PASS often assumes an ideal equal-power model, where power is evenly distributed across PAs on the same waveguide.
 However, according to a recent study \cite{wang2025modeling}, the proportional power model is more general in practice, which reveals that the power decays exponentially along the waveguide.
 Therefore, the first PA's position is of vital importance for antenna position optimization.
 
 Therefore, the original covert rate maximization problem \eqref{problem_2} can be converted into the following form:
 \begin{subequations} \label{problem_2a}
 	\begin{align}
 		\max_{\mathbf{x}_{\mathrm{init}},\tilde{\mathbf{w}}} \quad& \log _2\left( 1+P\gamma _{\mathrm{b}}\left( \mathbf{x}_{\mathrm{init}},\tilde{\mathbf{w}} \right) \right)  \label{eq:p2a-obj}\\
 		\mathrm{s}.\mathrm{t}.\quad &0\le [\mathbf{x}_{\mathrm{init}}]_n \le L-\Delta _x\left( M-1 \right) ,~~\forall n\in \mathcal{N} \label{eq:p2a-c1}\\
 		& P=\min \left\{ \frac{\Gamma _{\mathrm{w}}}{g_{\mathrm{w}}^{\max}\left( \mathbf{X},\tilde{\mathbf{w}} \right)},P_{\max} \right\}.
 		\label{eq:p2a-c2}
 	\end{align}
 \end{subequations}
Problem \eqref{problem_2a} is a non-convex optimization problem with highly-coupled variables, which is difficult to solve from the perspective of conventional optimization.
To solve this problem with low complexity, we adopt a twin particle swarm optimization (TwinPSO) approach to solve \eqref{problem_2a}.
The canonical PSO algorithm is a gradient-free, heuristic algorithm inspired by the foraging behavior of birds or fish swarms, which aims to secure the optimum solution over the whole feasible search space \cite{Kennedy1995particle, hu2024joint}. 
In contrast to the exhaustive search methods, PSO enables the information exchange among particles, allowing them to exploit both their individual best positions and the global best found by the swarm.
Therefore, based on this policy, the global optimal solution is expected to be found after several iterations.
Compared to deep learning methods, due to the gradient-free feature of PSO, it is more suitable for dealing with constraints that might hinder the back propagation step, such as \eqref{eq:p2a-c1} and \eqref{eq:p2a-c2}.

To handle multiple optimization variables, TwinPSO extends conventional PSO by assigning different roles to two types of particles: beamforming-seeking (BS) particles and position-seeking (PS) particles.
Then, the particle-level information exchange is allowed for particles in each swarm, whereas the swarm-level information exchange is allowed between different swarms.
The populations of the BS particles and the PS particles are denoted by $I$ and $J$, respectively.
Following this idea, in the $t$-th iteration, the status tuples of the $i$-th BS particle and the $j$-th PS particle are specified by
\begin{align}
	\varrho _{\mathrm{bs},i}^{\left( t \right)}&\triangleq \langle \mathbf{p}_{\mathrm{bs},i}^{\left( t \right)}\in \mathbb{C} ^{N\times 1},\mathbf{v}_{\mathrm{bs},i}^{\left( t \right)}\in \mathbb{C} ^{N\times 1},\dot{\mathbf{p}}_{\mathrm{bs},i}^{\left( t \right)}\in \mathbb{C} ^{N\times 1} \rangle, \notag 
	\\
	\varrho _{\mathrm{ps},j}^{\left( t \right)}&\triangleq \langle \mathbf{p}_{\mathrm{ps},j}^{\left( t \right)}\in \mathbb{R} ^{N\times 1},\mathbf{v}_{\mathrm{ps},j}^{\left( t \right)}\in \mathbb{R} ^{N\times 1},\dot{\mathbf{p}}_{\mathrm{ps},j}^{\left( t \right)}\in \mathbb{R} ^{N\times 1} \rangle, \notag 
\end{align}
where $\mathbf{p}_{\mathrm{bs},i}^{(t)}$ and $\mathbf{p}_{\mathrm{ps},j}^{(t)}$ represent the current position (or the value of the optimization variables, i.e., $\tilde{\mathbf{w}}$ and $\mathbf{x}_{\mathrm{init}}$) of the BS and PS particles, $\mathbf{v}_{\mathrm{bs},i}^{\left( t \right)}$ and $\mathbf{v}_{\mathrm{ps},j}^{\left( t \right)}$ denote their respective velocities, and $\dot{\mathbf{p}}_{\mathrm{bs},i}^{\left( t \right)}$ and $\dot{\mathbf{p}}_{\mathrm{ps},j}^{\left( t \right)}$ are the recorded current local best positions.
To meet the constraints \eqref{eq:p2a-c1}, the position of the $j$-th PS particle is processed by the following clamping function:
\begin{align}
	\left[ \mathbf{p}_{\mathrm{ps},j}^{(t)} \right]_n =
	\begin{cases}
		L^\prime, & \text{if } \left[ \mathbf{p}_{\mathrm{ps},j}^{(t)} \right]_n > 1, \\
		0, & \text{if } \left[ \mathbf{p}_{\mathrm{ps},j}^{(t)} \right]_n < 0, \\
		L^\prime \left[ \mathbf{p}_{\mathrm{ps},j}^{(t)} \right]_n, & \text{otherwise}.
	\end{cases} \label{eq:position_clamp}
\end{align}
where $n \in \mathcal{N}$ and $L^\prime \triangleq L-\Delta _x\left( M-1 \right)$.
Similarly, to satisfy the normalized power of $\tilde{\mathbf{w}}$, the position of the $i$-th BS particle is normalized via 
\begin{align}
	\mathbf{p}_{\mathrm{bs},i}^{\left( t \right)}=\mathbf{p}_{\mathrm{bs},i}^{\left( t \right)}/\left\| \mathbf{p}_{\mathrm{bs},i}^{\left( t \right)} \right\|. \label{eq:normalize}
\end{align}
\begin{algorithm}[t!]
	\SetAlgoLined
	\small
	\caption{TwinPSO Algorithm for SWSP-based Covert Communication}
	\label{alg:mwmp_teinpso}
	\KwIn{
		Precise position of Bob $\mathbf{r}_{\mathrm{b}}$; 
		Approximated position of Willie given by $(\bar{\mathbf{r}}_{\rm w}, \Delta_r)$; 
		Total power budget $P_{\max}$;
		Noise power uncertainty given by $(\bar{\sigma}_{\rm w}^2, \Delta_{\sigma})$; 
		Covertness requirement parameter $\rho_{\rm w}$;
		Initialized TwinPSO parameters $\omega _{\mathrm{bs}}$, $\omega _{\mathrm{ps}}$, $c_{\mathrm{bs},1}$, $c_{\mathrm{bs},2}$, $c_{\mathrm{ps},1}$, and $c_{\mathrm{ps},2}$;
		Max iteration $T$ and maximal velocity $v_{\max}$.
	}
	\KwOut{
		Optimized beamforming vector $\tilde{\mathbf{w}}$, 
		Optimized PA position vector $\mathbf{x}_{\rm{init}}$.
	}
	\DontPrintSemicolon
	\SetKwFunction{FReturn}{Return}
	
	\textbf{Initialization}: 
	Set iteration index $t \leftarrow 1$;
	Randomly initialize the positions of all the particles;
	For $\forall i, j$, initialize the best fitness scores by $R_{\mathrm{bs}, i}^{\mathrm{loc}}=R_{\mathrm{ps}, j}^{\mathrm{loc}}=R_{\mathrm{bs}}^{\mathrm{glo}}=R_{\mathrm{ps}}^{\mathrm{glo}}= -\infty$;
	Randomly initialize the global and local best positions of the particles.
	
	\For{$t \le T$}{
		\tcp{Position Optimization:}
		Obtain $\tilde{\mathbf{w}}$ by normalizing $\mathbf{p}_{\mathrm{bs}}^{\mathrm{glo}}$ via \eqref{eq:normalize}\;
		\For{$j=1,2,...,J$}{
			Clamp the position of the $j$-th PS particle via \eqref{eq:position_clamp}\;
			Compute the fitness score $R_{\mathrm{ps}, j}$ with fixed $\tilde{\mathbf{w}}$ via \eqref{eq:fitness_func}\;
			\If{$R_{\mathrm{ps}, j} > R_{\mathrm{ps}, j}^{\mathrm{loc}}$}{
				Update the local best fitness score $ R_{\mathrm{ps}, j}^{\mathrm{loc}}$\;
				Update the local best position $ \mathbf{p}_{\mathrm{ps}, j}^{\mathrm{loc}}$\;
				}
		}
		Set the best local position among the PS swarm as the global best position $\mathbf{p}_{\mathrm{ps}}^{\mathrm{glo}}$\;
		Update the velocities of the PS particles via \eqref{eq:up_ind_ps} \;
		Clamp these velocities to fit interval $[-v_{\max}, +v_{\max}]$\;
		Update the positions of the PS particles via \eqref{eq:ps_movement} \;
		Let $\mathbf{x}_{\mathrm{init}} \leftarrow \mathbf{p}_{\mathrm{ps}}^{\mathrm{glo}}$ as the optimized initial position of PAs\;
		\tcp{Beamforming Optimization:}
		\For{$i=1,2,...,I$}{
			Normalize the position of the $i$-th BS particle via \eqref{eq:normalize}\;
			Compute the fitness score $R_{\mathrm{bs}, i}$ with fixed $\mathbf{x}_{\mathrm{init}}$ via \eqref{eq:fitness_func}\;
			\If{$R_{\mathrm{bs}, i} > R_{\mathrm{bs}, i}^{\mathrm{loc}}$}{
				Update the local best fitness score $ R_{\mathrm{bs}, i}^{\mathrm{loc}}$\;
				Update the local best position $ \mathbf{p}_{\mathrm{bs}, i}^{\mathrm{loc}}$\;
				}
		}
		Set the best local position among the BS swarm as the global best position $\mathbf{p}_{\mathrm{bs}}^{\mathrm{glo}}$\;
		Update the velocities of the BS particles via \eqref{eq:up_ind_bs} \;
		Clamp these velocities to fit interval $[-v_{\max}, +v_{\max}]$ \;
		Increment index $t\leftarrow t+1$ \;
	}
	\Return{$\tilde{\mathbf{w}}$ and $\mathbf{x}_{\mathrm{init}}$}
\end{algorithm}
To evaluate the optimality of the positions of particles, the fitness function needs to be formulated by jointly considering \eqref{eq:p2a-obj} and \eqref{eq:p2a-c2}, which therefore yields:
\begin{align}
	R\left( \mathbf{p}_{\mathrm{ps}}, \mathbf{p}_{\mathrm{bs}} \right)
	= \log_2 \bigg( 1 + &\min \Big\{  \frac{\Gamma_{\mathrm{w}} \gamma_{\mathrm{b}}\left( \mathbf{p}_{\mathrm{ps}}, \mathbf{p}_{\mathrm{bs}} \right)}
	{g_{\mathrm{w}}^{\max}}, \notag \\
	&\qquad ~ P_{\max} \gamma_{\mathrm{b}}\left( \mathbf{p}_{\mathrm{ps}}, \mathbf{p}_{\mathrm{bs}} \right)
	\Big\} \bigg), \label{eq:fitness_func}
\end{align}
where $\mathbf{p}_{\mathrm{ps}}$ and $\mathbf{p}_{\mathrm{bs}}$ represent the positions of arbitrary PS and BS particles, respectively. 
After evaluation, the local best fitness scores are recorded as $R_{\mathrm{bs}, i}^{\mathrm{loc}}$ and $R_{\mathrm{ps}, j}^{\mathrm{loc}}$ for the $i$-th BS and the $j$-th PS particle, respectively.
Additionally, the corresponding best positions are recorded as $\mathbf{p}_{\mathrm{bs},i}^{\mathrm{loc}}$ and $\mathbf{p}_{\mathrm{ps},j}^{\mathrm{loc}}$, respectively.
Then, considering the entire particle populations, the global fitness scores for the BS and the PS swarm are respectively logged as $R_{\mathrm{bs}}^{\mathrm{glo}}$ and $R_{\mathrm{ps}}^{\mathrm{glo}}$, with the global best positions $\mathbf{p}_{\mathrm{bs}}^{\mathrm{glo}}$ and $\mathbf{p}_{\mathrm{ps}}^{\mathrm{glo}}$ being documented at the same time.
It is noted that the fitness function \eqref{eq:fitness_func} is determined by both the positions of BS and PS particles.
Therefore, in order to ensure a smooth optimization process, when the PS particles are examined, we fix the positions of the BS particles to the global best position in their swarm and vice versa.
This allows for inter-swarm information exchange.
Then, within one particular swarm, the following update rule is applied for the individual BS and PS particles, which are respectively given by 
\begin{align}
	\mathbf{v}_{\mathrm{bs},i}^{\left( t+1 \right)}&=\omega _{\mathrm{bs}}\mathbf{v}_{\mathrm{bs},i}^{\left( t \right)} +c_{\mathrm{bs},1}\theta _{\mathrm{bs},1}( \mathbf{p}_{\mathrm{bs},i}^{\mathrm{loc}}-\mathbf{p}_{\mathrm{bs},i}^{\left( t \right)} ) \notag \\ &\qquad \qquad \qquad \qquad +c_{\mathrm{bs},2}\theta _{\mathrm{bs},2}( \mathbf{p}_{\mathrm{bs}}^{\mathrm{glo}}-\mathbf{p}_{\mathrm{bs},i}^{\left( t \right)} ). \label{eq:up_ind_bs}
	\\
	\mathbf{v}_{\mathrm{ps},j}^{\left( t+1 \right)}&=\omega _{\mathrm{ps}}\mathbf{v}_{\mathrm{ps},j}^{\left( t \right)}+c_{\mathrm{ps},1}\theta _{\mathrm{ps},1}( \mathbf{p}_{\mathrm{ps},j}^{\mathrm{loc}}-\mathbf{p}_{\mathrm{ps},j}^{\left( t \right)} ) \notag \\ &\qquad \qquad \qquad \qquad  +c_{\mathrm{ps},2}\theta _{\mathrm{ps},2}( \mathbf{p}_{\mathrm{ps}}^{\mathrm{glo}}-\mathbf{p}_{\mathrm{ps},j}^{\left( t \right)} ) 
	, \label{eq:up_ind_ps}
\end{align}
where $\omega _{\mathrm{bs}}$ and $\omega _{\mathrm{ps}}$ are the inertia weights that balance exploitation and exploration; $c_{\mathrm{bs},1}$ and $c_{\mathrm{ps},1}$ are cognitive weights that control the importance of self-experiences;
$c_{\mathrm{bs},2}$ and $c_{\mathrm{ps},2}$ are the social weights that indicate the influence of the experience of the entire swarm;
$\theta _{\mathrm{bs},1}, \theta _{\mathrm{bs},2}, \theta _{\mathrm{ps},1}, \theta _{\mathrm{ps},2} \in [0, 1]$ are random variables that forces particles to explore.
It is noted that according to \eqref{eq:up_ind_bs} and \eqref{eq:up_ind_ps}, the inner-swarm information exchange is enabled by the global best position. 
Finally, the next position of the BS and PS particles can be calculated according to 
\begin{align}
	\mathbf{p}_{\mathrm{bs},i}^{\left( t+1 \right)}&=\mathbf{p}_{\mathrm{bs},i}^{\left( t \right)}+\mathbf{v}_{\mathrm{bs},i}^{\left( t+1 \right)}, \label{eq:bs_movement}
	\\
	\mathbf{p}_{\mathrm{ps},j}^{\left( t+1 \right)}&=\mathbf{p}_{\mathrm{ps},i}^{\left( t \right)}+\mathbf{v}_{\mathrm{ps},j}^{\left( t+1 \right)}.\label{eq:ps_movement}
\end{align}
The overall TwinPSO optimization algorithm is summarized by \textbf{Algorithm \ref{alg:mwmp_teinpso}}.

\subsection{Optimality and Complexity Analysis}  
\subsubsection{Optimality} Unlike the SWSP case, the TwinPSO algorithm cannot guarantee global optimality.
However, by employing a larger population of particles, a near-optimal solution is more likely to be attained, where an optimality-complexity trade-off is introduced. 

\subsubsection{Complexity} 
Given the complexity of evaluating the fitness function as $C$, the computational complexity for updating one particle can be specified by 
$\mathcal{O}(C+N)$ with $N$ being the number of waveguides (or dimensional of each particle's position).
Then, considering the particle populations for the PS and BS are $J$ and $I$, the computational complexity for one iteration is $\mathcal{O} (\left( J+I \right) \left( C+N \right) )$.
Jointly considering the maximal iteration $T$, the overall complexity can be computed by $\mathcal{O} (T\left( J+I \right) \left( C+N \right) )$.

\begin{figure}[t!]
	\centering
	\begin{subfigure}{0.47\linewidth}
		\includegraphics[width=\textwidth]{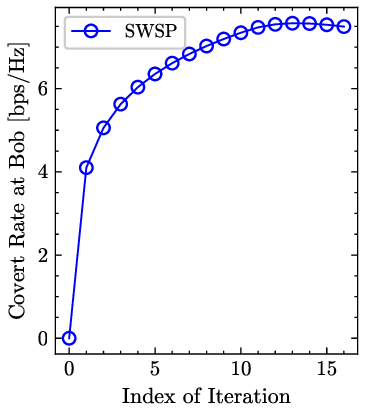}
		\caption{Convergence behavior of 1D search algorithm.}
		\label{fig:swsp-convergence}
	\end{subfigure}
	\begin{subfigure}{0.5\linewidth}
		\includegraphics[width=\textwidth]{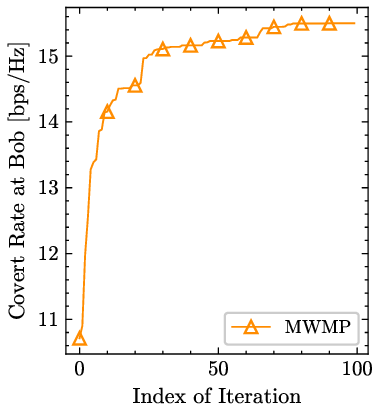}
		\caption{Convergence behavior of Twin PSO algorithm.}
		\label{fig:mwmp-convergence}
	\end{subfigure}
	\caption{The convergence behavior of the proposed algorithms.}
	\label{fig:convergence_behavior}
\end{figure}

\section{Numerical Results}\label{sect:results}
 In this section, we present the numerical results to verify the effectiveness of the proposed methods.
 The following parameter setups are utilized throughout this section, unless otherwise specified.

 For propagation-related parameters, the carrier frequency $f$ and the total power budget $P_{\max}$ are respectively set to 28 GHz and $30$ dBm, while the effective refractive index is set to $n_{\mathrm{eff}}=1.4$ according to \cite{wang2025modeling, ding2024flexible,ouyang2025array}.
 For the legitimate user Bob, the noise power is set to $\sigma_{\mathrm{b}}=-100~\mathrm{dBm}$.
 At the malicious user Willie, the noise uncertainty model is specified by $\bar{\sigma}_{\mathrm{w}}^2 = -70~{\mathrm{dBm}}$ and $\Delta_{\sigma, \mathrm{dB}}=2~{\mathrm{dB}}$.
 For the physical layout of PASS, the heights of waveguides are uniformly set to $h=3~{\mathrm{m}}$, and the length of the waveguides is set to $L=25~\mathrm{m}$.
 In the SWSP case, the waveguide is placed along the $x$-axis.
 In the MWMP case, $N=4$ waveguides are uniformly placed paralleled to the $x$-axis with a separation interval of $\Delta_{\rm wg}=3~\mathrm{m}$;
 besides, the number of PAs amounted on each waveguide is set to $M=3$ with a fixed interval of $\Delta_{x}=\frac{\lambda_{\rm c}}{2}$.
 For the hyperparameters of the algorithms, the search resolution for the 1D linear search in the SWSP case is set to $\Delta_{P}=P_{\max}/10^5$.
 In the MWMP case, the hyperparameters for the TwinPSO approach are listed as $I=J=30$, $\omega _{\mathrm{bs}}=\omega _{\mathrm{ps}}=0.8$, $c_{\mathrm{bs},1}=c_{\mathrm{bs},2}=2.0$, and $c_{\mathrm{ps},1}=c_{\mathrm{ps},2}=2.0$, with maximal iteration times $T=100$ and maximal allowed particle velocity $v_{\max}=0.3$.
 To perform Monte Carlo simulations, there are $\# 200$ samples that are uniformly distributed over a region specified by $x \times y = [0~\mathrm{m}, 25~\mathrm{m}] \times [-7.5~\mathrm{m}, 7.5~\mathrm{m}]$.
 The number of samples in the position uncertainty region of Willie is set to $|\mathcal{C}|=5$.

In order to showcase the effectiveness of the proposed algorithms, the following benchmarks are considered throughout the simulation.
\begin{itemize}
	\item \textbf{MIMO ZF}: This benchmark employs the conventional MIMO technology with an $N$ half-wavelength uniform linear array (ULA). The center of the ULA is placed at the origin of the coordinate system. Zero-forcing (ZF) beamforming, i.e., putting the exact location of Willie into the null space, is utilized based on a fully-digital antenna configuration.
	The transmit power is set to ensure the beam gain at Willie cannot exceed the threshold $\Gamma_{\mathrm{w}}$.
	\item \textbf{MIMO MRT}: This benchmark also utilizes ULA for transmitting, with the same power control policy as the ``MIMO ZF" setup. 
	Different from the former benchmark, maximum ratio transmission (MRT) is utilized to align the beamforming vector with the channel vector of Bob.
	The power control policy is identical to the ``MIMO ZF" setup.
	\item \textbf{PASS ZF}: This benchmark utilizes MWMP PASS for transmission. However, instead of performing dedicated PA position optimization, the initial PAs on the waveguides are placed closest to Bob's position.
	The beamforming policy is set to zero forcing.
	\item \textbf{PASS MRT}: This benchmark uses MWMP PASS as well. Its only difference from the ``PASS ZF" configuration is that MRT is used as a beamforming policy.
\end{itemize}
 \begin{figure}[t!]
	\centering
	\begin{subfigure}{0.72\linewidth}
		\includegraphics[width=\textwidth]{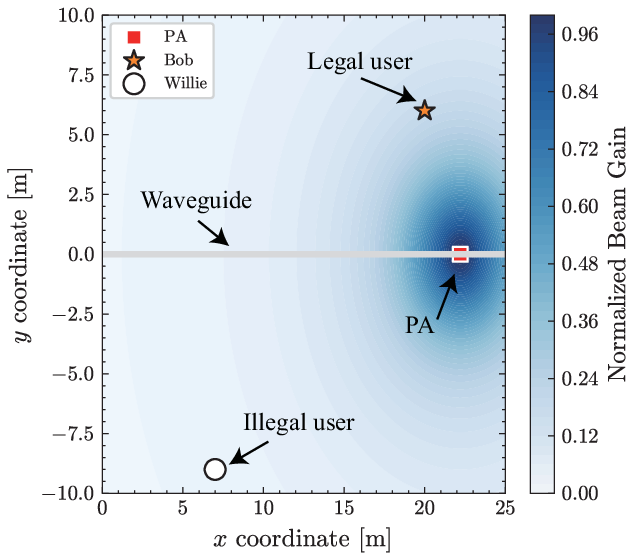}
		\caption{SWSP Case}
		\label{fig:swsp-pattern}
	\end{subfigure} \\
	\begin{subfigure}{0.72\linewidth}
		\includegraphics[width=\textwidth]{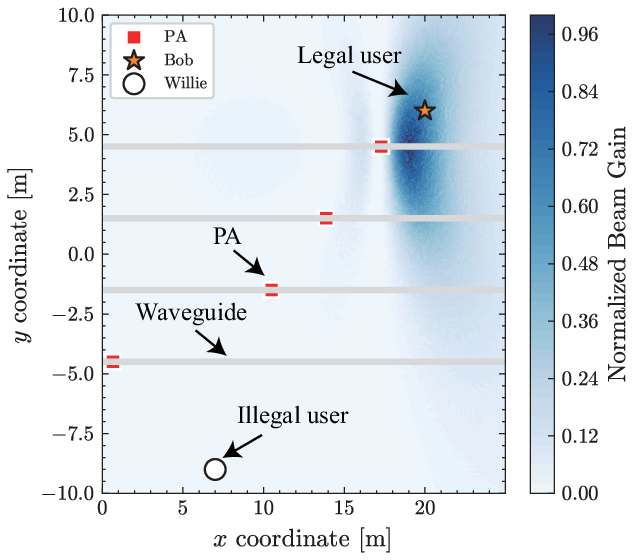}
		\caption{MWMP Case}
		\label{fig:mwmp-pattern}
	\end{subfigure}
	\caption{The obtained beam patterns for both the SWSP and the MWMP case. Beam gains are normalized with respect to their respective maximum value.}
	\label{fig:patterns}
\end{figure}

\subsection{Convergence Behaviors of Proposed Algorithms}
To evaluate the effectiveness of the proposed algorithms, the convergence behaviors of the proposed algorithms under the same Bob and Willie layout are illustrated in Fig. \ref{fig:convergence_behavior}.
The results are obtained through averaging $\# 200$ random initializations.
As observed, the 1D search algorithm for the SWSP case in Fig. \ref{fig:swsp-convergence} takes $\# 15$ iterations to converge, indicating that a quick and low-complexity convergence is achieved by integrating the closed-form solution for the optimal PA position.
In comparison, the TwinPSO approach for the MWMP case in Fig. \ref{fig:mwmp-convergence} takes more iterations and converges after $\# 60$ iterations.
Although the TwinPSO approach needs more iterations to converge, it can attain a high-quality suboptimal solution with fewer iterations.
Moreover, due to the adoption of multiple PAs, the MWMP case can achieve a much higher covert rate than that obtained by the SWSP case as a result of enhanced antenna flexibility.

\subsection{Illustration of Optimized Beam Patterns}
In Fig. \ref{fig:patterns}, the optimized beam patterns for both the SWSP and MWMP cases are showcased for the same user layout, i.e., $\mathbf{r}_{\rm b}=[20~m, 6~m, 0~m]^{\mathrm{T}}$ and $\bar{\mathbf{r}}_{\rm w}=[7~m, -9~m, 0~m]^{\mathrm{T}}$.
For the SWSP case shown in Fig. \ref{fig:swsp-pattern}, the position of the PA is placed at $[22~m, 0~m]$, which is closer to the position of Bob.
Consequently, the received transmit power at Bob is much larger than that of malicious Willie, thereby ensuring the covert transmission requirement.
By tuning the position of PA, the PASS can achieve a trade-off between ``maximizing Bob's covert rate" and ``reducing Willie's received power".
However, limited by the waveguide and single PA in the SWSP case, the high beam gain region can only be allocated beneath PA's position.
On the contrary, for the MWMP case shown in Fig. \ref{fig:mwmp-pattern}, the high beam gain region can be allocated above the position of Bob, as a result of the broader coverage of PASS and pinching beamforming.
Hence, the covert rate can be significantly enhanced according to the results in Fig. \ref{fig:convergence_behavior}.
Moreover, due to the imitated number of antennas equipped by each waveguide, beam can only be separated upon the region containing Bob rather than being focused at Bob's exact location. 

\subsection{Throughput Versus Target Total Error Rate}
\begin{figure}[t!]
	\centering
	\includegraphics[width=0.83\linewidth]{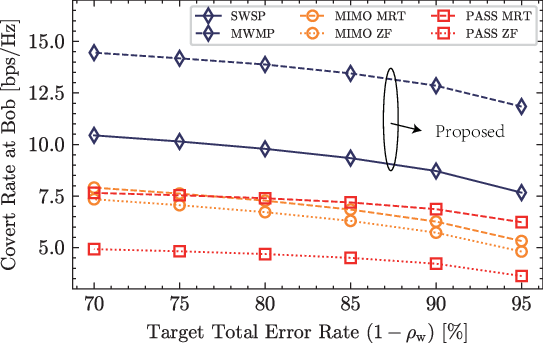}
	\caption{Illustrate of the covert rate at Bob versus the target total error rate $1-\rho_{\mathrm{w}}$.} \label{fig:covert_rate_vs_total_error_rate}
\end{figure}
Fig. \ref{fig:covert_rate_vs_total_error_rate} shows how the covert rate varies with the target total error rate at Willie, i.e., $1-\rho_{\mathrm{w}}$, which reflects the system's covertness requirement. 
In particular, a higher target total error rate implies more detection errors at the malicious user, indicating better covertness. 
As the figure illustrates, the covert rate at Bob decreases as the target total error rate increases. 
This is because enhancing covertness requires suppressing the beam gain at Willie, thereby enforcing Alice to reduce transmit power and ultimately diminish the covert rate at Bob.

Thanks to a more flexible antenna configuration, the MWMP PASS achieves higher covert rates than the SWSP scenario. 
Compared to conventional MIMO benchmarks such as ``MIMO ZF" and ``MIMO MRT," PASS offers significantly better covert performance, highlighting the benefits of pinching beamforming. 
Additionally, PASS reduces implementation costs, as even a single waveguide with a single PA can achieve a higher covert rate than any other benchmarks with multiple antennas. 
Compared to heuristic PASS optimization schemes i.e., ``PASS ZF" and ``PASS MRT," the results show that dedicated pinching beamforming algorithms are essential to fully realize the potential of PASS.

\subsection{Throughput Versus Total Power Budget}
\begin{figure}[t!]
	\centering
	\includegraphics[width=0.83\linewidth]{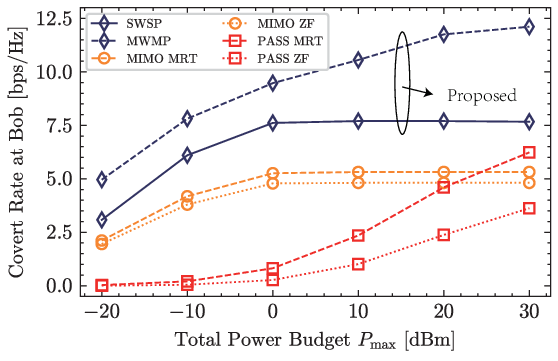}
	\caption{Illustrate of the covert rate at Bob versus the total power budget $P_{\max}$.} \label{fig:covert_rate_vs_transmit_power}
\end{figure}
Fig. \ref{fig:covert_rate_vs_transmit_power} illustrates the impact of the total power budget on the covert rate.
As $P_{\max}$ increases, the covert rate increases accordingly, but with gradually diminishing returns.
This observation is due to the covertness constraint (beam gain threshold) at Willie, which will prevent Alice from transmitting with full available power.
The superior performance of PASS is confirmed by the performance gain over the conventional MIMO technologies. 
Moreover, the poor performance of heuristic optimizations in the ``PASS ZF'' and ``PASS MRT''  benchmarks highlights that the PASS performance cannot be ensured unless dedicated optimization techniques are employed.

\subsection{Throughput Versus Radius of Willie's Position Uncertainty Region}
\begin{figure}[t!]
	\centering
	\includegraphics[width=0.83\linewidth]{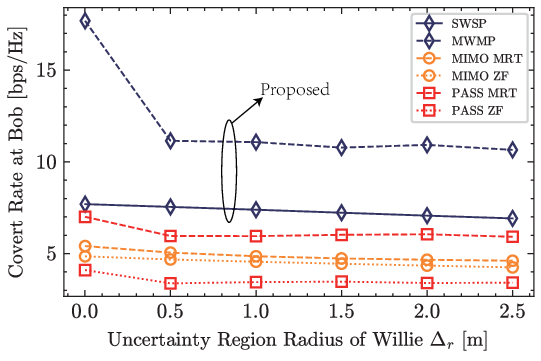}
	\caption{Illustrate of the covert rate at Bob versus the radius of Willie's position uncertainty region $\Delta_r$.} \label{fig:covert_rate_vs_uncertainty_rad}
\end{figure}
Fig. \ref{fig:covert_rate_vs_uncertainty_rad} examines how changes in the radius of Willie’s position uncertainty region, denoted by $\Delta_r$, affect the covert rate.
As shown in this figure, when the $\Delta_r$ increases, a drop in Bob's covert rate can be observed.
Furthermore, it can also be observed that the MWMP PASS setups, including ``MWMP proposed'', ``PASS MRT'', and ``PASS ZF", experience a more significant performance loss at the small $\Delta_r$ region.
This is due to the sampling method used for addressing the covertness constraint.
In particular, the optimization process targets the sample within Willie's uncertainty region that receives the highest beam gain, which represents the worst-case scenario.
Consequently, when the exact Willie's position is utilized for testing, there would be a performance loss.
In contrast, this performance loss is less pronounced for the SWSP setup, as the closed-form optimal position of PA is derived, thereby eliminating the need for the sampling method.

\section{Conclusion}\label{sect:conlusion}
This paper investigated a covert communication system with the SWSP and the MWMP PASS under the uncertainty of Willie's position and noise.
For the SWSP case, the closed-form optimal PA position was derived.
Building on this, a 1D linear search algorithm was utilized to maximize the transmission rate while ensuring covertness.
For the MWMP case, the TwinPSO approach was proposed to perform pinching beamforming, providing performance gains through enhanced flexibility.
The numerical results showcased that the effectiveness of the proposed method and also demonstrated the advantages of PASS over conventional antenna architectures.
As this paper demonstrates the superiority of PASS, future research can focus on exploring its flexibility in handling more challenging scenarios, such as moving malicious eavesdroppers, multiple collaborative eavesdroppers, or the use of PASS to emit jamming signals.

\begin{appendices}
    \section{Proof of Lemma \ref{lemma:forbidden_zone}} \label{forbidden_zone_proof}
\begin{figure}[!t]
		\centering
		\includegraphics[width=0.7\linewidth]{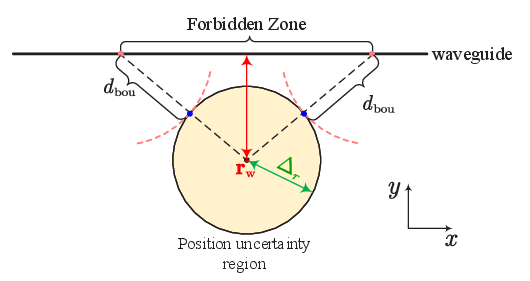}
		\caption{Illustration of the forbidden zone of PA under the SWSP case.} \label{fig:appendix_1}
	\end{figure}

    For each possible position of Willie, denoted by $\mathbf{r}_{\mathrm{w}}$, Alice can derive the minimal total error rate according to \textbf{Lemma \ref{lemma:opt_detection}}, which can be expressed as
    \begin{align}
        \xi =\frac{1}{2\ln (\Delta _{\sigma} )}\ln \left( \frac{\Delta _{\sigma}^{2}\bar{\sigma}_{\mathrm{w}}^{2}}{\bar{\sigma}_{\mathrm{w}}^{2}+\Delta _{\sigma}P\frac{\eta}{\left\| \mathbf{r}_{\mathrm{w}}-\mathbf{p} \right\| _{2}^{2}}} \right). \tag{A-1} \label{eq:a-1}
    \end{align}
    For notation simplicity, we denote the distance between PA and a possible position of Willie as $d_{\mathrm{pw}}\triangleq \left\| {\mathbf{r}}_{\mathrm{w}}-\mathbf{p} \right\|$.
    When \eqref{eq:p1-c3} is satisfied, we have $\xi \ge 1-\rho _{\mathrm{w}} $.
    Then, jointly considering $\Delta_{\sigma} >1$ and $0 \le \rho _{\mathrm{w}} \le 1$, mathematical manipulations can lead us to the following equation:
    \begin{align}
		d_{\mathrm{pw}}^{}\ge \sqrt{P\eta \Delta _{\sigma}/\left( \bar{\sigma}_{\mathrm{w}}^{2}\left( \Delta _{\sigma}^{2\rho _{\mathrm{w}}}-1 \right) \right)}\triangleq d_{\mathrm{bou}}, \tag{A-2} \label{eq:A-2}
    \end{align}
	where $d_{\mathrm{bou}}$ is defined as the minimum distance to guarantee covertness.
	Given the height of the waveguide is fixed, we project \eqref{eq:A-2} onto the $XOY$ plane according to
	\begin{align}
		\left( d_{\mathrm{pw}}^{\prime} \right) ^2\triangleq \left( x_{\mathrm{w}}-x_{\mathrm{p}} \right) ^2+\left( y_{\mathrm{w}}-y_{\mathrm{p}} \right) ^2\ge d_{\mathrm{bou}}^{2}-h^2\triangleq \left( d_{\mathrm{bou}}^{\prime} \right) ^2 \tag{A-3} \label{eq:A-3}
	\end{align}
	It is noted that when $d_{\mathrm{bou}} <h $, \eqref{eq:A-3} can always be satisfied, thereby the covertness constraint \eqref{eq:p1-c3} can be discarded.
	Otherwise, when $d_{\mathrm{bou}} \ge h $, the distance relationship between Willie and PA must be discussed.
	We define \emph{Forbidden Zone} as a distance region within which the covertness constraint is not always satisfied across Willie's position uncertainty region.
	To determine the boundaries of this zone, we consider PA's movement along the waveguide and iteratively construct circles of $d_{\mathrm{bou}}^\prime$.
	If any point within the position uncertainty region falls into the constructed circles, the covertness constraint will be violated.
	Consequently, the boundaries of the forbidden zone exist when the constructed circle intersects with the circle representing the position uncertainty region of Willie.
	This process can be visualized in Fig. \ref{fig:appendix_1}.
   	For notation simplicity, we denote $d_{\mathrm{tot}}\triangleq d_{\mathrm{bou}}^{\prime}+\Delta_{r}$.
   	Hence, in light of Fig. \ref{fig:appendix_1}, when $|y_{\rm w}| < d_{\mathrm{tot}}$, the forbidden zone can be expressed as
   	\begin{align}
   		&\theta =\mathrm{arc}\cos \left\{ \frac{\left| y_{\mathrm{w}} \right|}{d_{\mathrm{tot}}} \right\},
   		\tag{A-4a} \label{eq:A-4a} \\
   		&x_{\mathrm{w}}-d_{\mathrm{tot}}\sin \left( \theta \right) \le x \le x_{\mathrm{w}}+d_{\mathrm{tot}}\sin \left( \theta \right). \tag{A-4b} \label{eq:A-4b}
   	\end{align} 
   	Otherwise, there is no forbidden zone for any possible location of Willie.
   	Here, the forbidden region can be computed.
   	
   	\section{Proof of Theorem \ref{theorem:optimal_transmit_power}}\label{optimal_transmit_power}
   Given the unit power beamforming vector $\tilde{\mathbf{w}}$ fixed, the original beamforming vector for PASS can be expressed as  $\mathbf{w}=\sqrt{P}\tilde{\mathbf{w}}$ with $\left\| \tilde{\mathbf{w}} \right\|=1$.
   	Therefore, according to \eqref{eq:beam_gain} and letting $g_{\mathrm{w}}^{\max}\left( \mathbf{X},\tilde{\mathbf{w}} \right) ={\max }_{\mathbf{r}_{\mathrm{w}}\in \mathcal{G}}~g_{\mathrm{w}}^{}\left( \mathbf{X},\tilde{\mathbf{w}} \right)$, the beam gain at Willie can be computed by
   	\begin{align}
		P\le \Gamma _{\mathrm{w}}/g_{\mathrm{w}}^{\max}\left( \mathbf{X},\tilde{\mathbf{w}} \right) \triangleq g(\tilde{\mathbf{w}}). \tag{B-1} \label{eq:B-1} \notag 
   	\end{align}
   	To prove this theorem that the optimal transmit power is attained at the boundary of \eqref{eq:B-1}, i.e., $P^{\mathrm{opt}}=g(\tilde{\mathbf{w}})$, we adopt of the proof by contradiction.
    Specifically, we assume that the optimal solution is not attained at the boundary;
    in the sequel, we show that a better solution can be constructed and satisfied all constraints, thus challenging the optimality of the original solution;
   	thereby, the optimal solution must be achieved at the boundary. 
   	
    Following this logic flow, if $\exists~P^{\mathrm{opt}} < g(\tilde{\mathbf{w}})$ is the optimal solution to the optimization problem, a better solution $P^\prime$, satisfying $P^\prime \le g(\tilde{\mathbf{w}})$, can be established by $P^\prime = P^{\mathrm{opt}}+ \delta$ for an arbitrary infinitesimal positive value $\delta>0$.
   	According to \eqref{eq:p2-obj}, the covert rate at Bob is a monotonically increasing function with respect to $P$.
   	Thus, the covert rate achieved using $P^\prime$ is larger than that using $P^{\mathrm{opt}}$.
   	Consequently, the optimality of the initial solution does not hold anymore.
	Thus, this proof is completed here.
\end{appendices}

\bibliographystyle{IEEEtran}
\bibliography{IEEEabrv, mybib}
\end{document}